\documentclass[final,5p,times,twocolumn]{elsarticle}

\usepackage{lscape}
\usepackage{amsmath}
\usepackage{xspace} 
\usepackage{hyperref}
\usepackage{dirtytalk}
\usepackage{graphicx}
\usepackage{subfig}
\usepackage{multirow}
\usepackage{lipsum}
\usepackage{booktabs}
\usepackage{natbib}
\usepackage{longtable}
\usepackage{caption}
\usepackage{adjustbox}
\usepackage{tabularx}
\usepackage{floatrow} 
\usepackage[vlines]{tabularht}
\usepackage{pbox}
\usepackage{pifont}% http://ctan.org/pkg/pifont
\newcommand*\colourcheck[1]{%
  \expandafter\newcommand\csname #1check\endcsname{\textcolor{#1}{\ding{52}}}%
}
\colourcheck{blue}
\colourcheck{green}
\colourcheck{red}
\usepackage{float}
\floatstyle{plaintop}
\restylefloat{table}
\makeatletter
\newcommand{\thickhline}{%
    \noalign {\ifnum 0=`}\fi \hrule height 1pt
    \futurelet \reserved@a \@xhline
}
\newcolumntype{"}{@{\vrule width 1pt}}
\makeatother
\newlength{\Oldarrayrulewidth}

\usepackage[linesnumbered,ruled,vlined]{algorithm2e}
\SetKwInput{KwInput}{Input}                % Set the Input
\SetKwInput{KwOutput}{Output}              % set the Output
\usepackage[table,xcdraw]{xcolor}

\newsavebox\MBox

\biboptions{sort&compress}
\journal{XX}

\begin{document}
\begin{frontmatter}
\title{InfLocNet: Enhanced Lung Infection Localization and Disease Detection from Chest X-Ray Images Using Lightweight Deep Learning}

\author[label1]{Md. Asiful Islam Miah\corref{cor1}}
\address[label1]{ Southeast University, Dhaka, Bangladesh}
\cortext[cor1]{I am corresponding author}
\ead{asiful.miah@seu.edu.bd}
\author[label2]{Shourin Paul}
\ead{shourinpaul@gmail.com}
\author[label2]{Sunanda Das}
\ead{sunanda@cse.kuet.ac.bd}
\author[label2]{M. M. A. Hashem}
\ead{hashem@cse.kuet.ac.bd}
\address[label2]{Khulna University of Engineering \& Technology (KUET), Khulna, Bangladesh}

\begin{abstract}

In recent years, the integration of deep learning techniques into medical imaging has revolutionized the diagnosis and treatment of lung diseases, particularly in the context of COVID-19 and pneumonia. This paper presents a novel, lightweight deep-learning-based segmentation-classification network designed to enhance the detection and localization of lung infections using chest X-ray (CXR) images. By leveraging the power of transfer learning with pre-trained VGG-16 weights, our model achieves robust performance even with limited training data. The architecture incorporates refined skip connections within the UNet++ framework, reducing semantic gaps and improving precision in segmentation tasks. Additionally, a classification module is integrated at the end of the encoder block, enabling simultaneous classification and segmentation. This dual functionality enhances the model's versatility, providing comprehensive diagnostic insights while optimizing computational efficiency. Experimental results demonstrate that our proposed lightweight network outperforms existing methods in terms of accuracy and computational requirements, making it a viable solution for real-time and resource-constrained medical imaging applications. Furthermore, the streamlined design facilitates easier hyperparameter tuning and deployment on edge devices, broadening the model's applicability across various domains. This work underscores the potential of advanced deep learning architectures in improving clinical outcomes through precise and efficient medical image analysis. Our model achieved remarkable results with an Intersection over Union (IoU) of 93.59\% and a Dice Similarity Coefficient (DSC) of 97.61\% in lung area segmentation, and an IoU of 97.67\% and a DSC of 87.61\% for infection region localization. Additionally, it demonstrated high accuracy of 93.86\% and sensitivity of 89.55\% in detecting chest diseases, highlighting its efficacy and reliability.

\end{abstract}

\end{frontmatter}

%%%%%%%%%%%%%%%  ChestInf-Net       %%%%%%%%%%%%%%%%%%%%%%%%%%%%%%

\section{Introduction}
\label{sec:introduction}

%%%%%%%%%%%%%%%%%%%%%%%%%%%%%%%%%%%%%%%%%%%%%%%%%%%%%%%% Start %%%%%%%%%%%%%%%%%%%%%%%%%%%%%%%%%%%%%%%%%

Respiratory diseases constitute a formidable global health challenge, affecting millions of individuals annually with significant morbidity and mortality rates. Ranging from chronic ailments like chronic obstructive pulmonary disease (COPD) and lung cancer to acute infections such as pneumonia and COVID-19, these conditions encompass a broad spectrum of pathologies, necessitating comprehensive approaches to diagnosis, treatment, and management. At the center of this intricate physiological system are the lungs, vital organs responsible for facilitating the exchange of oxygen and carbon dioxide essential for sustaining life. The intricate interplay of cellular and molecular processes within the lungs ensures the delivery of oxygen to tissues and organs while expelling waste gases, thus maintaining homeostasis and overall health.

Despite the crucial role of the lungs in respiratory function, they are susceptible to various diseases influenced by multifactorial determinants. Environmental exposures, including air pollution \cite{lungorg_climate_change}, tobacco smoke \cite{hosny2018artificial}, occupational hazards  \cite{hassaballah2020deep}, and indoor pollutants  \cite{tahamtan2020real}, pose significant risks to lung health. Moreover, genetic predispositions, infections, and lifestyle choices further compound the vulnerability of the lungs to disease \cite{xia2020evaluation}. Notably, smoking remains a primary risk factor for developing lung diseases, with tobacco smoke containing numerous harmful chemicals that can inflict damage on lung tissue and increase susceptibility to conditions like COPD and lung cancer \cite{who_top_causes_death}.

\begin{figure*}[t]
    \centering
    \includegraphics[width=\textwidth, height = 4in]{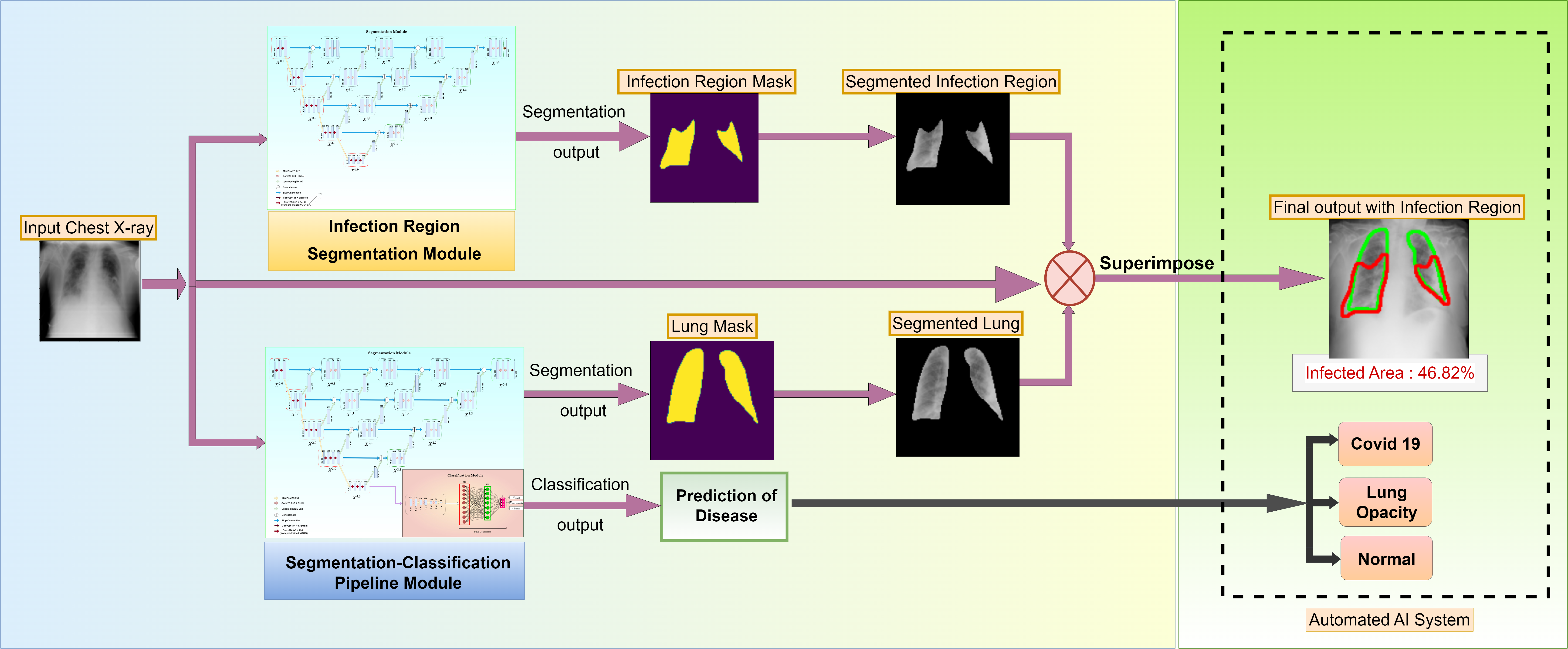}
    \caption{Proposed Network for Chest Disease Detection and Infection Region Localization.}
    \label{prop_arch}
\end{figure*}

Among the myriad respiratory diseases, pneumonia stands as a prominent public health concern, exerting a substantial toll on global health systems and economies. Characterized by infectious inflammation of the lungs, pneumonia results from the infiltration of alveoli and bronchioles by bacterial, viral, fungal, or parasitic pathogens, leading to inflammation and fluid accumulation \cite{who_pneumonia}. Its clinical manifestations span a wide spectrum, ranging from mild respiratory symptoms to life-threatening complications, particularly in vulnerable populations such as infants and older adults. According to the World Health Organization (WHO), pneumonia accounts for a significant proportion of global fatalities, with children under the age of five bearing a disproportionate burden of morbidity and mortality \cite{rudan2008epidemiology}.

Despite advances in medical diagnostics and therapeutics, pneumonia diagnosis remains a complex endeavor, often requiring a multifaceted approach encompassing clinical evaluation, imaging studies, and laboratory tests \cite{cherian2005standardized}. However, access to these diagnostic modalities remains limited in resource-constrained settings, hindering timely and accurate identification of the disease. Furthermore, the interpretation of radiological imaging, such as chest X-rays (CXRs) and computed tomography (CT) scans, relies heavily on the expertise of radiologists, whose availability may be constrained, particularly in underserved regions \cite{kalra2015ct}.

In recent years, the global community has grappled with the unprecedented challenge posed by the COVID-19 pandemic, caused by the novel coronavirus SARS-CoV-2. COVID-19 presents with a spectrum of clinical manifestations, ranging from mild flu-like symptoms to severe respiratory distress and multi-organ failure. Its rapid transmission dynamics, coupled with a high degree of variability in symptomatology and disease severity, have overwhelmed healthcare systems worldwide, necessitating innovative approaches to disease detection, management, and containment \cite{guan2020zhong}.

Previous research indicates that interpreting chest X-ray (CXR) images for lung diseases can be difficult due to their broad spectrum of features. However, characteristic ground-glass opacities (GGOs) in COVID-19 patients present as distinct patches in the lungs, complicating the density and size of lesions over time. Furthermore, consolidations in lung tissue, seen in viral pneumonia and COVID-19, add complexity to CXR interpretation. Despite their value, CXR findings vary with infection severity, making accurate diagnosis challenging. Thus, the proposed two-stage framework combines classification and segmentation to enhance lung disease detection and infection localization.
Recent advancements in medical imaging underscore the potential of chest radiography, particularly chest X-ray (CXR) images, for diagnosing pneumonia and COVID-19 pneumonia. Despite their widespread availability, CXR images have limited sensitivity in detecting chest diseases, prompting exploration into advanced computational and deep learning methods. Studies by Souid et al. \cite{souid2021classification}, Goram et al. \cite{alshmrani2023deep}, and Pooja et al. \cite{yadav2021lung} have achieved high accuracies in disease classification, with values around 94.5\% and F1 scores up to 95.62\% across various lung diseases. However, challenges persist in effectively diagnosing infections within segmented lung regions, as highlighted in studies by Tabik et al. \cite{tabik2020covidgr} and Robert Hertel et al.  \cite{hertel2022deep}, who did not focus on chest infections and employed two separate models for classification and segmentation, potentially leading to computational inefficiencies and inadequate resolution of diagnostic challenges within segmented lung regions.
Based on the effective performance of discrete deep learning models in both classification and segmentation tasks, this research introduces an integrated framework for identifying lung diseases and segmenting infections using chest X-ray (CXR) images. The proposed framework employs a new classification-segmentation pipeline that leverages pretrained classification and segmentation networks. Functioning as a unified model, it not only categorizes candidate images into specific classes such as COVID-19, viral pneumonia, and normal, but also conducts infection segmentation in pneumonia and COVID-19 positive images. Developed based on the pretrained VGG16 model, this framework is labeled as ChestInf-Net.
This paper presents a novel segmentation-classification network, leveraging deep learning techniques, to effectively detect lung diseases and localize infection areas. To enhance model performance and robustness, we employed transfer learning with pre-trained VGG-16 weights from the ImageNet dataset, which accelerates learning and adapts well to limited training data\cite{simonyan2014very}. Furthermore, we refined the skip connections in the UNet++\cite{unetpp} architecture to minimize semantic gaps and improve precision\cite{zhou2020unet++}. Additionally, our approach includes a classification module to classify diseases based on input from the pre-trained encoder, alongside an infection region localization method using segmentation masks. Notably, the proposed architecture is designed with fewer parameters, ensuring resilience against overfitting and establishing it as a lightweight network\cite{howard2017mobilenets}.  Detailed network architecture and output generation procedures are elucidated in Sections \ref{Methodology} and \ref{Experiment}, respectively. Figure \ref{prop_arch} shows an abstract depiction of our proposed network.

This paper introduces a novel deep learning-based segmentation-classification network tailored for chest X-ray images. The key contributions of our study include:

\begin{itemize}

   \item \textbf{Comprehensive Framework:} This study introduces an integrated framework for lung disease detection and infection segmentation in chest X-ray (CXR) images, providing a unified solution for medical image analysis.
   \item \textbf{Unified Architectural Ingenuity:} Unlike conventional methodologies, the proposed segmentation-classification network amalgamates these critical tasks into a singular, harmonized model, facilitating seamless segmentation of chest images and disease prognostication with unparalleled efficiency.
   \item \textbf{Optimized Model Design:} The architecture of the proposed network is optimized for efficiency, featuring a lightweight design with reduced parameters while maintaining high accuracy levels in segmentation and disease classification tasks.
   \item \textbf{Elevated Performance Pinnacle:} Experimental evaluations demonstrate the superior performance of the proposed model compared to existing methods, showcasing its effectiveness in accurately detecting lung diseases and segmenting infections from CXR images.
   \item \textbf{Precision-Enhanced Infection Assessment:} The incorporation of an infection module within the proposed framework enables more precise quantification of infected regions, particularly in identifying areas affected by COVID-19 or pneumonia, thereby enhancing diagnostic capabilities for clinicians.
    % \item To design a complete segmentation model.
\end{itemize}

%%%%%%%%%%%%%%%%%%%%%%%%%%%%%%%%%%%%%%%%%%%%%%%%%%%%%%%%%% End %%%%%%%%%%%%%%%%%%%%%%%%%%%%%%%%%%%%%%%%

\section{Literature Review}
\label{BACKGROUND}
%%%%%%%%%%%%%%%%%%%%%%%%%%%%%%%%%%%%%%% Start   %%%%%%%%%%%%%%%%%%%%%%%%%%%%%%%%%%%%%%%%%%%%%%%%%%%
Chest radiography is a popular medical imaging technique for diagnosing and identifying pneumonia and other lung disorders. As previously discussed, deep learning models have been successfully employed in screening, diagnosing, and treating chest diseases using chest CT and CXR images. However, if we look at prior literature and studies, we can see that researchers prefer to use chest X-ray (CXR) images over CT scans since CXR images are more readily available from numerous places. In line with this observation, this section reviews selected literature and identifies research gaps through a comprehensive analysis of deep learning models used for chest disease detection and infection segmentation from CXR images.

Recent research has shown promising results in detecting underlying features from radiography pictures for diagnostic analysis utilizing cutting-edge computational and deep learning methods. Souid et al.\cite{souid2021classification} developed a deep learning model using MobileNet V2 to classify lung diseases from chest X-rays. The study achieved an accuracy of about 94.5\% in classifying chest diseases. Their dataset has almost 14 different binary classes of disease. Goram et al.\cite{alshmrani2023deep} proposed a deep-learning architecture for multiclass lung disease detection. The study used a convolutional neural network (CNN) model to identify the most prevalent chest diseases, such as pneumonia, tuberculosis, and lung cancer. The study classified chest illnesses with an F1 score of 95.62\%. Pooja et al.\cite{yadav2021lung} proposed an unsupervised framework as it is hard to find a huge number of labeled data for a novel disease. They trained and tested their framework on 6 large publicly available lung disease datasets and obtained an accuracy of 94\% - 99.5\%. Recently  Maider Abad et al.\cite{abad2024generalizable} have developed a disease detection system using model ensemble on CXR images.  The research utilized 26,047 images sourced from six different datasets to refine three pre-trained models: IRV2, ResNet50, and DenseNet121.The proposed ensemble method achieved an accuracy of 97.38\% and an AUC of 97.35\% on the internal validation set. On the external validation set, the ensemble model outperformed individual models, with an accuracy of 81.16\%, precision of 77.11\% and sensitivity of 80.97\%.

\begin{table*}
    \centering
    \caption{A brief summary of related works}
    \begin{adjustbox}{width=\columnwidth,center}
    \begin{tabular}{m{10em}m{15em}m{10em}m{15em}} 
    \hline
        \textbf{Authors} & \textbf{Techniques} & \textbf{Dataset} & \textbf{Weakness/Remarks} \\ \hline
        % Ronenberger et al.\cite{unet} & U-Net & ISBI 2012 & It has Semantic gap problem and deconvolution layer. \\ \hline
        % Badrinarayanan et al. \cite{segnet} & U-Net - deconvolution + upsampling - skip connection &  CamVid road scene segmentation and SUN RGB-D indoor scene segmentation & It has no skip connection. This led to lower performance of the model. \\ \hline
        % Amrita et al.\cite{kaur2021ga} & U-Net + Genetic Algorithm & BRAT18,19, Shenzhen, and 3D-IRCADb-01 & Dont have any prediction module. It has large complexity.\\ \hline
        Anas M. Tahir et al. \cite{tahir2021covid} & U-Net  & COVID-QU-Ex Dataset & Can misclassify covid with pneumononia.\\ \hline
        % Zhou et al. \cite{unet++} & Nested skip connection & cell nuclei, colon polyp, liver, and lung nodule & It is very complex because of using nested skip connection. \\ \hline
        % Huimin et al. \cite{unet3+} & Full scale skip connection + CGM module &  ISBI LiTS 2017 and self build & It only reduce the false positive outputs, not the false negative outputs. \\ \hline
        % Chen et al. \cite{dualchexnet} & DenseNet Block + ResNet Block & ChestX-ray14 & This architecture works well for chest X-ray dataset, computationally expensive and time consuming \\ \hline
        % Al-Masni et al. \cite{cmm} & Contextual multi-scale multi-level network + inversion recovery &  ISIC 2017, DRIVE , and BraTS 2018 & Has great performance in multimodal biomedical images. \\ \hline
        % Kwon et al. \cite{pggan} & Weighted multi-scale similarity & NIH Chest X-ray & Domain specific architecture. \\ \hline

        S.Tabik et al. \cite{tabik2020covidgr} & Segmentation Based Cropping, COVID-SDNet & COVIDGR-10 & Preprocessing based on cropping the image \\ \hline
        
        Tarun Agrawal et al.\cite{agrawal2023covid}  & UNet with attention mechanism and  a convolution-based atrous spatial pyramid pooling module  & QaTa-COV19 dataset \cite{degerli2021covid}  & lower lesion segmentation performance and no disease detection is performed \\ \hline

         Degerli et al.\cite{degerli2021covid}  & collaborative human–machine approach  & QaTa-COV19 dataset \cite{qatacov19_dataset}  & Can detect and localize infection only for COVID \\ \hline

          N.B. Prakash et al.\cite{prakash2021deep}  & Extended the SqueezeNet classification model with Grad-CAM and super pixel pooling  &  Kaggle CXR public database \cite{chowdhury2020covid}  & Lack of generalizability due to imbalance dataset. No precise boundary for the infection region \\ \hline

        Robert hertel et al. \cite{hertel2022deep} & ResUNet & Chest Radiography Datase, RICORD dataset & Can't predict the infection percentage. \\ \hline
        % UNet3+ & 8,963,073 \\ \hline
        % UNet3+ & 8,963,073 \\ \hline
    \end{tabular}
    \end{adjustbox}
    \label{tab:rel_works}
\end{table*}

To do the detection tasks many researchers prioritize segmentation tasks before undertaking detection tasks.  By isolating and focusing on the area of interest, such as lung segmentation in chest X-ray images, detection algorithms can analyze relevant features more effectively, leading to improved precision in identifying abnormalities or diseases within the segmented region. In this context, Tabik et al.\cite{tabik2020covidgr} devised a segmentation-classification model aimed at detecting COVID-19. In the segmentation phase, instead of conducting actual segmentation, they opted to crop the image, focusing solely on the lung area while eliminating extraneous sections. However, this approach may not entirely eradicate irrelevant data from the image. Robert Hertel et al.\cite{hertel2022deep} devised a deep learning-based system for segmentation and classification to identify COVID-19. However, their approach did not involve addressing chest infections, and they employed separate models for classification and segmentation, resulting in high computational costs. the researchers achieved a Dice Similarity Coefficient (DSC) of 0.95, indicative of strong performance. Despite this, their disease detection accuracy, at 84\%, did not meet desired standards. However, detection through segmented lung images has some disadvantages. This approach poses a risk of information loss, as subtle disease indicators outside the lung region may be inadvertently removed during segmentation. Though it may boost accuracy within the confines of the model's specifications, this approach may inadvertently diminish its generalizability across broader datasets or real-world scenarios. Additionally, the sequential nature of lung segmentation followed by disease detection prolongs processing time, limiting the applicability of the system in real-time scenarios. Lastly, the accuracy of disease detection heavily relies on the precision of lung segmentation, meaning any inaccuracies in this process directly affect the system's ability to accurately identify diseases.

There is a limited number of studies that have specifically concentrated on infection region localization. However, several notable research efforts exist in this area.
Anas M. Tahir et al. \cite{tahir2021covid} has done an excellent job by developing a U-Net-based architecture for COVID-19 identification and infection localization. The authors propose a systematic and unified approach for lung segmentation and COVID-19 localization with infection quantification from chest X-ray (CXR) images. They constructed a large benchmark dataset with 33,920 CXR images, including 11,956 COVID-19 samples, and performed extensive experiments using state-of-the-art segmentation networks.  They have utilized two U-Net architectures in their proposed system: one to generate the entire lung mask from CXR images and the other to produce the mask for the infected portion of the lung. Subsequently, these generated masks are superimposed onto the CXR image to localize and quantify COVID-19-infected lung regions. The generated infection mask is then used to detect COVID-19. They attained an Intersection over Union (IOU) and Dice Similarity Coefficient (DSC) of around 83.05\% and 88.21\%, respectively, for infection region segmentation, which is noteworthy although not optimal. The detection mechanism focuses on detecting the infected part, although it may have difficulty separating COVID-19 from cases of viral pneumonia. Furthermore, their infection region quantification is based on identifying the percentage of infected areas, without an evaluation method to account for potential mismeasurements in circumstances where the projected infection mask gives false positive or negative readings.

Degerli et al.\cite{degerli2021covid}  introduced a novel method for generating COVID-19 infection maps. They utilized a substantial dataset comprising approximately 120,000 chest X-ray images, which included 2951 samples of COVID-19. Furthermore, they publicly released the dataset along with ground-truth segmentation masks for COVID-19. The study achieved high sensitivity levels of 98.37\% and a specificity of 99.16\%, indicating a low false alarm rate. For infection localization, their best-performing network has achieved an F1 score of 85.81\%. According to their huge dataset, they have a limited number of COVID-19 images which can affect their model's generalization capability. However, their proposed method is solely focused on localizing COVID-19 infections. Hence, there is undoubtedly potential for enhancement, especially concerning both localizing and quantifying infection regions. This might entail computing the overall percentage of lung area affected by infection while also assessing the presence of false positive or false negative outcomes. Such an approach would assist medical professionals in quantifying severity and tracking the progression of chest diseases.

N.B. Prakash et al. \cite{prakash2021deep} have also made significant contributions to chest disease detection. They have developed a deep learning model specifically for COVID-19 detection and have further identified infection regions utilizing transfer learning in conjunction with superpixel-based segmentations. The model achieves excellent accuracy in binary and multi-class classifications, with the binary classifier scoring 99.53\% and the multi-class classifier scoring 99.79\%. The COVID-SSNet also uses superpixel segmentation of activation maps to isolate areas of interest, which improves the diagnostic usefulness of chest X-ray images for COVID-19 treatment. However, the dataset which is utilized to train and evaluate their model is quite small. It includes 219 COVID-19-positive images, 1345 viral pneumonia, and 1341 normal images. The dataset's imbalance, where COVID-19-positive images constitute only about one-sixth of the other classes,  may jeopardize the model's generalizability. Their methodology extends the basic SqueezeNet classification model by combining Grad-CAM and superpixel pooling methods. Grad-CAM is applied to the final convolutional layer to generate an activation map for the complete CXR image. These activation maps are then passed to the superpixel pooling layer and segmented to highlight the most important features, which are then superimposed onto the original CXR image.  However, this method may not be optimal for precisely diagnosing infection regions since it might show vulnerable areas that are not part of the lungs. Furthermore, the network's capability is limited to providing only a coarse localization, so this approach may fail to measure infection rates or precisely identify infected regions inside the lung.

Tarun Agrawal et al.\cite{agrawal2023covid} also contribute to this pandemic by developing a UNet-based encoder-decoder architecture for COVID-19 lesion segmentation. To boost performance, the proposed model includes an attention mechanism as well as a convolution-based atrous spatial pyramid pooling module. The suggested model produced values of 0.8325 and 0.7132 for the dice similarity coefficient and the Jaccard index, respectively. But their work doesn't have any detection system as doctors have to predict it by seeing the lesion. Also,  if we compare it with some recent works, we can see that the lesion segmentation performance is comparatively low. They also mentioned some cases where their model failed to segment the lesion properly and according to them the reason for the failure cases may be the presence of rib cages and clavicle bone in CXR images. Perhaps the implementation of some proper preprocessing techniques could enhance the segmentation model's efficacy.

%%%%%%%%%%%%%%%%%%%%%%%%%%%%%%%%%%%%%%%%%%%%%%%%%%%%%%%%%%    End    %%%%%%%%%%%%%%%%%%%%%%%%%%%%%%%%%%%%%%%%

\section{Motivation and high level consideration}
\label{MOTIVATION}

%%%%%%%%%%%%%%%%%%%%%%%%%%%%%%%%%%%%%%% Start   %%%%%%%%%%%%%%%%%%%%%%%%%%%%%%%%%%%%%%%%%%%%%%%%%%%

\subsection{Efficient Training with Transfer Learning}
\label{subsec:trans}
\begin{figure}[h!]
    \centering
    \includegraphics[ scale=0.3]{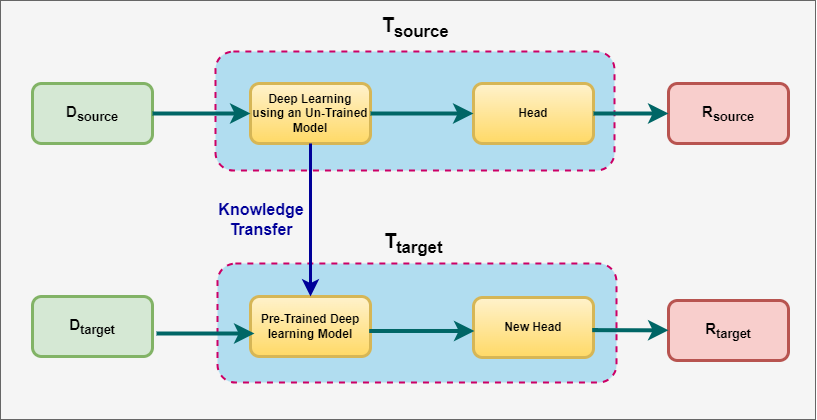}
    \caption{Transfer learning technique.}
    \label{trans_learn}
\end{figure}

Large quantities of labeled data are frequently needed for the construction of machine learning models from scratch, but obtaining this data can be challenging and expensive. Transfer learning addresses this challenge by leveraging knowledge from one task to enhance performance on a related task. In this method, a model is initially trained on a substantial dataset for a base task. The pre-trained model is then repurposed with a new head for the target task, facilitating faster and more robust learning. Formally, if $D_{source}$ and $T_{source}$ represent the base domain and task, and $D_{target}$ and $T_{target}$ represent the target domain and task, transfer learning aims to improve the learning of the target predictive result, $R_{target}$ in $T_{target}$, by utilizing knowledge from $D_{source}$ and $T_{source}$ \cite{transfer}. This approach enhances model robustness and accelerates learning by using pre-existing knowledge. In our case, transfer learning was employed to extract primitive features such as edges and structures from a large image dataset, which then initialized and improved the learning process for biomedical image segmentation. Figure \ref{trans_learn} illustrates this process, where a pre-trained model is adapted with a new head to address specific, related tasks, reducing the need for extensive labeled data and accelerating the learning process, particularly useful for medical image segmentation.

\subsection{Streamlining Skip Path Dense Convolutional Blocks}
\label{subsec:res}

In U-Net, U-Net++, and similar architectures, some useful information gets lost during the max-pooling operations. The U-Net++ architecture is enriched with skip connections. While UNet++ enhances medical image segmentation by refining skip pathways and adding extra convolutional layers, it introduces several disadvantages. The increased complexity and computational requirements make the model resource-intensive, leading to longer training times and higher memory consumption. Additionally, the intricate design of UNet++ complicates hyperparameter optimization and reduces model interpretability, crucial for clinical applications. Moreover, the model's complexity may lead to overfitting, particularly with small datasets, requiring extensive regularization techniques. To address these issues, we propose reducing the dense convolutional blocks within the skip paths of UNet++. By minimizing these blocks, several benefits are realized, including a reduction in computational demands and memory usage, faster training speeds, and decreased risk of overfitting. Furthermore, this optimization simplifies hyperparameter tuning, enhances model interpretability, and facilitates deployment on edge devices. Overall, this adjustment optimizes the model's efficiency while maintaining its performance, making it suitable for resource-constrained or real-time scenarios.

\subsection{Classification of Disease using Classification Module}
\label{subsec:cgm}

In computer vision, integrating an image classification module into a segmentation model enhances the model's capability to provide detailed insights into the contents of an image. By classifying objects or regions into pre-defined categories, the classification module enables the segmentation model to make more informed decisions about how to segment the image, leading to a comprehensive understanding of its contents.

The classification module comprises additional convolutional layers, pooling operations, and fully connected layers to extract relevant features and make predictions about the input image. During training, it is optimized alongside the segmentation component using a combined loss function, ensuring effective learning of both tasks. During inference, the classification module utilizes the encoded representation to predict classes or categories within the image, augmenting segmentation results for comprehensive understanding and interpretation of detected features.

There are certain advantages of including a classification module to the encoder block of an encoder-decoder based segmentation model. Firstly, it enables simultaneous performance of segmentation and classification tasks, enhancing the model's versatility. Furthermore, the classification module facilitates identification of specific classes or categories within segmented regions, particularly beneficial in medical imaging for accurate diagnosis and treatment planning. Additionally, integrating the classification module optimizes feature extraction, reducing redundant computations, and enhancing overall efficiency, thereby broadening the model's applicability across diverse domains.

\subsection{Finding Infected Regions and Calculate the Severity and Exactness of the Infection in Lungs}

It's crucial to accurately determine the severity of lung infections, as they can lead to serious health issues if left untreated. A combination of the lung mask and the infection mask provides a visual representation of the infected regions, enabling medical professionals to assess the extent of the infection and determine the necessary course of action. The use of these two masks in the evaluation process is essential for making informed decisions about the patient's health, ensuring that appropriate treatment is provided in a timely manner, and monitoring the progress of the patient over time. This information is crucial for providing high-quality medical care and ensuring positive health outcomes.

The Lung Segmentation-Classification model generates a lung mask that outlines the lung region in an image. The Infection Region Segmentation module, on the other hand, identifies the infected regions within the lung mask. The combination of these two masks provides a comprehensive picture of the severity of the lung infection, as it outlines both the lung region and the extent of the infection within it. The severity of infection was calculated using the Eq. \ref{eq:1}.

The exactness of the infection localization was evaluated using Eq \ref{eq:8}, which calculates the Infection Intersection over Union. It measures the overlap between the predicted infection region and the actual infection region, relative to the union of both regions.
\small
\begin{equation} \label{eq:8}
\scriptsize
   Infection\;IoU = \frac{\mid{{Actual\;Infection\;Area}\cap {Predicted\;Infection\;Area}}\mid}{\mid{{Actual\;Infection\;Area}}\mid \cup \mid{{Predicted\;Infection\;Area}}\mid}
\end{equation}

\begin{equation} \label{eq:16}
{Actual\;Infection\;Area} = \frac{\sum {ground\; Inf\; Mask}}{\sum {ground\; Whole\; Mask}}
\end{equation}

\begin{equation} \label{eq:1}
Predicted\;Infection\;Area = \frac{\sum Y_{infected\;Region}}{\sum Y_{whole\;Lung}} * 100\%
\end{equation}

where, \\ 
ground Inf Mask = Ground Truth mask of infected region\\
ground Whole Mask = Ground Truth mask of whole lung \\
Y\textsubscript{infected Region} = Predicted mask of infected region\\
Y\textsubscript{whole Lung} = Predicted mask of whole lung \\

By evaluating the severity of infection in the lungs, medical professionals can make informed decisions on treatment and monitor the progress of the patient.

%%%%%%%%%%%%%%%%%%%%%%%%%%%%%%%%%%%%%%% End   %%%%%%%%%%%%%%%%%%%%%%%%%%%%%%%%%%%%%%%%%%%%%%%%%%%

\begin{figure*}[hbt!]
 \centering
{\includegraphics[width=\textwidth]{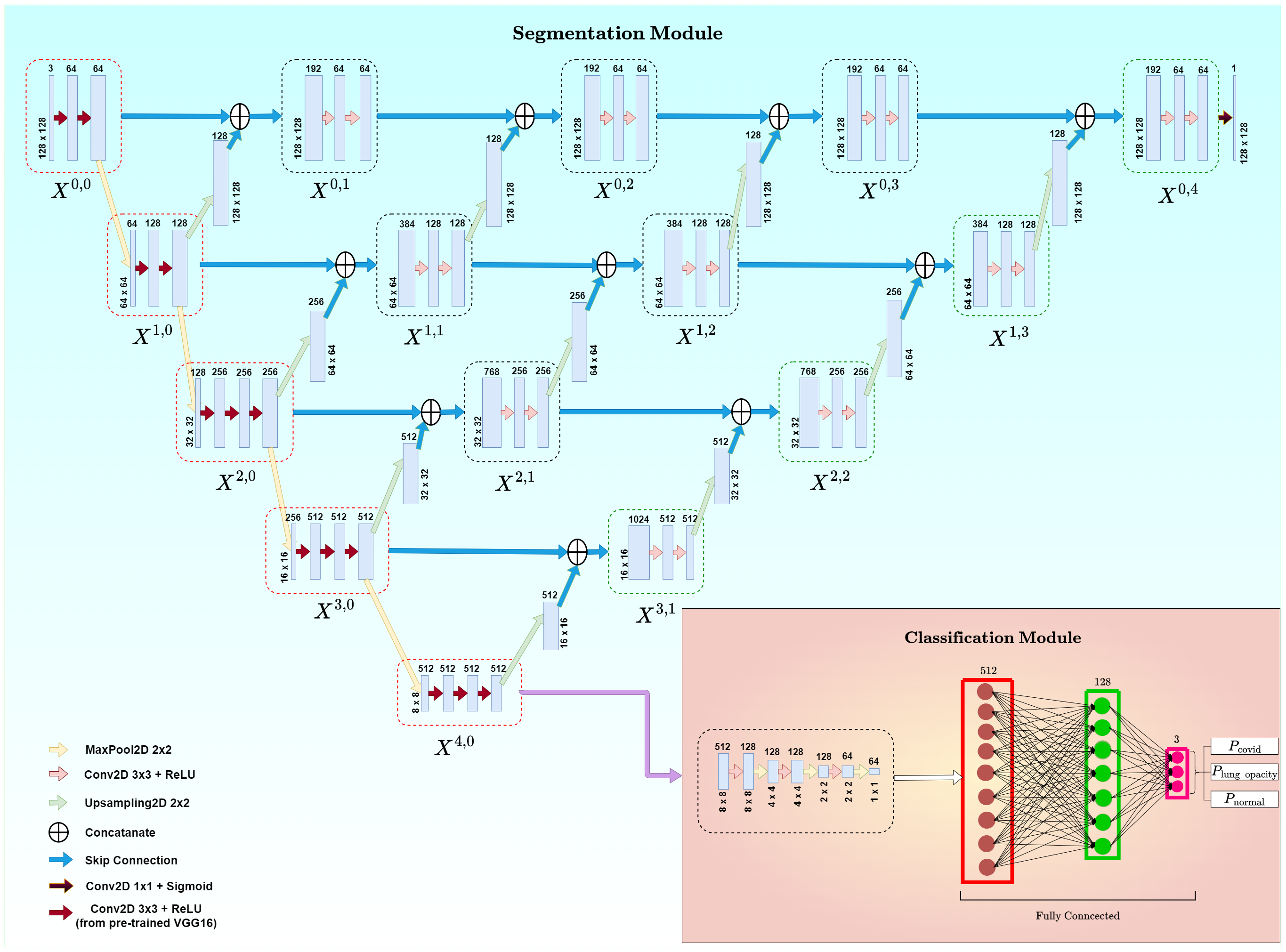}}
  \caption{Lung Segmentation and Classification Module Architecture}
  \label{diagram3}
\end{figure*}

\section{Proposed Architecture}
\label{Methodology}

%  ..............
%  Chest InfNet......Start..................
%  ........................

In response to the limitations of current state-of-the-art networks, we introduce Chest-InfNet, a novel framework aimed at addressing the challenges of lung disease detection and infection region segmentation. This framework is designed as an integrated model capable of performing both classification and segmentation tasks. It consists of two subnetworks where one handles segmentation and classification jointly, while the other focuses exclusively on infection region segmentation. Notably, our proposed networks have fewer parameters approximately 18 million and 15 million, respectively compared to U-Net and U-Net++, which have around 28 million and 37 million parameters, respectively. This section provides an overview of the implementation of our proposed system, including its underlying architectures and methodologies. Figure \ref{prop_arch} illustrates the overall structure of the proposed system.

\subsection{Segmentation-Classification Pipeline Network Architecture}

The primary objective of Chest-InfNet is to segment out lung regions before proceeding to classify the disease type using extracted features. Lung region contains crucial portions of disease-related information. Figure \ref{diagram3} outlines the proposed architecture for the Segmentation-Classification pipeline network, delineating the precise workflow between segmentation and classification tasks. This proactive method ensures an accurate evaluation of chest X-ray images, where segmentation provides a strong basis for precise disease identification. Through this integrated framework, Chest-InfNet ensures a precise and efficient solution for automated disease detection and classification in medical imaging. 

The segmentation module for segmentation task consists of encoder, central block, decoder and modified skip connection.

\subsubsection{Pre-trained Encoder using Transfer Learning}

% \begin{figure}[h!]
%     \centering
%     \includegraphics[width=.4\textwidth]{trans_learn_vgg16.png}
%     \caption{The proposed pre-trained encoder.}
%     \label{pre_enc}
% \end{figure}

The encoder of our model comprises four blocks. Each encoder block contains a convolutional block derived from the VGG16 pretrained model, serving as the backbone architecture. Each encoder block includes convolutional block followed by a max-pooling layer, which aids in downsampling the feature maps to capture more abstract representations from the input image. 
We used the VGG-16 network to learn the base task $T_S$ from the ImageNet dataset\cite{imagenet} as a pretrained model. Though the task of VGG-16 network is classification, it can help a segmentation network on the encoder phase. We used the VGG-16 network because it is not complex and more similar to the encoder part of the U-Net based segmentation network. We derived the first five convolutional layers from the base network because these layers are responsible for recognizing the primitive features of an image. We used the pre-trained weights as the initialization for the encoder part of the proposed model.

\subsubsection{Center Block}

This block feeds data into the network's decoder sections through the incorporation of a convolutional block from the VGG-16 model. It essentially functions as a bridge to connect the two network segments. Moreover, this block serves as the basis for the classification module.

\subsubsection{Streamlining Skip Path}

The skip connection has been modified in order to minimize the computational cost and memory usage that were mentioned in sub-section \ref{subsec:res}. The encoder and decoder subnetworks are now more connected because of redesigned skip pathways. The feature maps of the encoder are received directly by the decoder in U-Net, but in UNet++, they pass through a dense convolution block and the number of blocks varies depending on the pyramid level.

In our proposed model, the skip pathway incorporates a concatenation layer before each convolution layer. This concatenation layer merges the output from the previous convolution layer within the same block and the up-sampled output from the lower block. This process aims to align the semantic level of the encoder feature maps with that of the feature maps in the decoder. The hypothesis is that when the received encoder feature maps and the associated decoder feature maps are semantically comparable, the optimizer will have a better accuracy of solving the optimization problem. The skip pathway can be expressed formally as follows: let $x^{i,j}$ represent the output of node $X^{i,j}$, where i denotes the encoder's down-sampling layer and j the convolutional block's along the skip pathway. The stack of feature
maps represented by $x^{i,j}$ is computed as

 \begin{equation} \label{eq:15}
x^{i, j}= \begin{cases}\mathcal{H}\left(x^{i-1, j}\right), & j=0, i>0 \\ \mathcal{H}\left(\left[x^{i, j-1}, \mathcal{U}\left(x^{i+1, j-1}\right)\right]\right), & j>0\end{cases}
\end{equation}
 
 where U(·) represents an up-sampling layer, [ ] represents the concatenation layer, and function H(·) is a two consecutive convolution operation followed by an activation function.
In essence, nodes at level $j > 0$ receive two inputs—one from the encoder sub-network and the other from the up-sampled output of the lower skip pathway—while nodes at level $i = 0$ receive only one input from the pre-trained layer of the encoder and nodes at levels $j = 0$ and $i > 0$ receive only one input from the previous layer of the encoder. We employ a dense convolution block along each skip pathway, and this is the reason how all previous feature maps accumulate and reach the present node.

\subsubsection{Decoder}

The decoder module is comprised of four decoder blocks. Each decoder block is preceded by an upsampling layer utilizing bilinear interpolation. Inside each block, there are two $3\times3$ convolution layers with ReLU activation. Feature maps from the convolution layers of the modified skip connections are concatenated with the corresponding decoder feature maps. Output from final layer of the decoder block is passed into a $1\times1$ convolution block and sigmoid activation to generate the predicted output segmentation mask. In binary classification issues, the sigmoid activation function is widely used to generate output values ranging from 0 to 1. This function is used to generate the output segmentation mask by classifying each pixel of the input image into two classes.

\subsubsection{Classification Module}

As mentioned in sub-section \ref{subsec:cgm}, the proposed model has a Classification module to handle the probable inherent classification task. 

The classification module within the Segmentation-Classification pipeline network utilizes deep features extracted by the encoder blocks. Initially, feature maps of size 8×8×512, obtained from the central block $X^{4,0}$, are inputted into the classification module. These feature maps undergo a series of transformations to enhance their discriminative power. First, they are resized to 1×1×64 dimensions, followed by three 3×3 convolutional layers with ReLU activation and 2×2 max-pooling layers to extract and amplify important features. Subsequently, the feature maps are flattened into a 1-dimensional vector for efficient processing. This vector is then fed through two dense layers, each comprising 512 and 128 neurons, respectively. Dropout layers are applied after each dense layer to mitigate overfitting and enhance the model's generalization capability. Finally, a dense layer with 3 neurons is employed to produce the probability distribution for each class. The class with the highest predicted probability is determined as the final classification result, providing the type of disease.

\subsection{Infection Region Segmentation Network Architecture}

The fundamental objective of the Chest-InfNet network is to effectively identify and localize the infection areas within the lung from chest X-ray images. This requires meticulously analyzing the imaging data to locate areas having infection-related characteristics such as opacities. For this purpose, we propose the Infection Segmentation Network, which exclusively utilizes the segmentation module. To identify the infection mask image from a chest x-ray in order to extract the infection area using the infection mask, the proposed Infection Segmentation Network uses Segmentation Module from the Segmentation-Classification Pipeline Network, excluding the Classification Module. The network can assist in the diagnosis of respiratory disorders, such as pneumonia or viral infections like COVID-19, by precisely locating these regions.

%  \begin{figure*}[hbt!]
%  \centering
% {\includegraphics[width=5in]{conf/proposed_inf.drawio.png}}
%   \caption{ Infection Region Segmentation Module Architecture.}
%   \label{diagram4}
% \end{figure*}

%  ..............
%  Chest InfNet.......End.................
%  ........................

%  ..............
%  Chest InfNet.......Start.................
%  ........................

\section{Experiment}
\label{Experiment}

% \usepackage{multirow}
\begin{table*}
 \centering
    \caption{Summary of the datasets used in the experiments.}
    \label{tab:dataset}
    
\begin{tabular}{ccccc} 

\hline
Modality & Dataset & \multicolumn{3}{l}{Total no. of images and it’s variations} \\ 
\hline

\multirow{7}{*}{Chest X-Ray} & \multirow{3}{*}{COVID-QU-Ex Dataset} & \multicolumn{1}{c}{\multirow{3}{*}{33000}} & \multicolumn{2}{c}{11955 (COVID-19)} \\ 
 &  & \multicolumn{1}{c}{} & \multicolumn{2}{c}{11261 (Pneumonia)} \\ 
 &  & \multicolumn{1}{c}{} & \multicolumn{2}{c}{10704 (Normal)} \\ \cline{2-5} 
 & \multirow{4}{*}{COVID-19 Radiography Dataset} & \multicolumn{1}{c}{\multirow{4}{*}{21165}} & \multicolumn{2}{c}{3616 (COVID-19)} \\  
 &  & \multicolumn{1}{c}{} & \multicolumn{2}{c}{10192 (Normal)} \\ 
 &  & \multicolumn{1}{c}{} & \multicolumn{2}{c}{6012 (Lung Opacity)} \\ 
 &  & \multicolumn{1}{c}{} & \multicolumn{2}{c}{1345 (Pneumonia)} \\ \hline
\end{tabular}
\end{table*}

\subsection{Datasets Acquisition}
\label{sec:data}

There are several difficulties in creating datasets for medical imaging, such as privacy concerns, the need for expert annotation, complex image acquisition processes, and the high costs of imaging equipment. In the domain of medical imaging, there are limited publicly accessible benchmark datasets, and each dataset only has a limited quantity of images. To evaluate the performance of our proposed network in segmentation and classification tasks, we selected two distinct Chest X-ray datasets, both openly accessible and accompanied by corresponding ground truths. Our experiments focus specifically on 2D images within each dataset. Below, we provide brief descriptions of these datasets and also table \ref{tab:dataset} provides a summary of their main attributes.

\subsubsection{COVID-QU-Ex Dataset\cite{tahir2021covid} }

\begin{figure}[hbt!]
 \centering
{\includegraphics[scale=0.4]{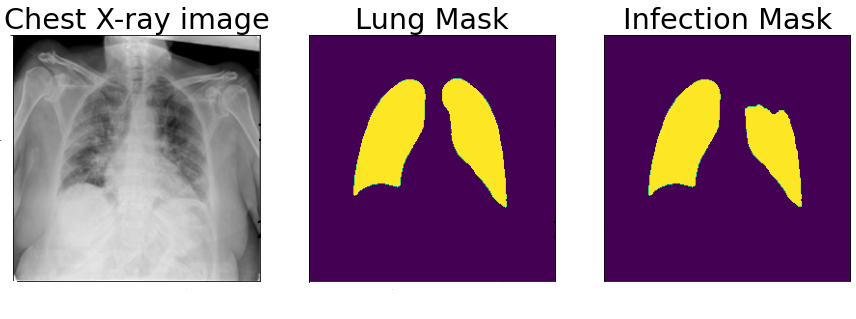}}
  \caption{
  Sample Chest X-ray Image and its Corresponding Masks of the COVID-QU-Ex Dataset}
  \label{diagram2}
\end{figure}

The COVID-QU-Ex dataset is a comprehensive repository of medical imaging data pertaining to the COVID-19 pandemic. Chest X-rays from patients with COVID-19 diagnoses  as well as from healthy people are included in this dataset. This dataset has been extensively used by researchers and was created especially for research and algorithmic development regarding COVID-19 detection and diagnosis through medical imaging. The dataset comprises a total of 33,000 Chest X-ray images, each accompanied by binary infection masks indicating three distinct classes: 11,955 Covid-19 cases, 11,261 instances of Non-Covid Viral or Bacterial Pneumonia infections, and 10,704 Normal cases. Fig. \ref{diagram2} visually depicts a sample from this dataset.

\subsubsection{COVID-19 Radiography Dataset\cite{chowdhury2020can} }

The COVID-19 Radiography dataset comprises chest X-ray images obtained from individuals diagnosed with COVID-19, as well as those from healthy subjects, serving as controls. This publicly available dataset is primarily used by researchers to train and evaluate different deep learning models. It is mainly intended for the development and evaluation of deep learning algorithms developed for COVID-19 and lung disease detection through chest X-ray images. The dataset helps the development of effective algorithms for early lung disease detection and diagnosis because it contains a significant number of annotated images that demonstrate the presence or absence of lung disease. This dataset contains a total of 21,165 images, accompanied by their  corresponding binary lung mask images. The dataset encompasses four classes, including 3,616 Covid-19 positive cases, 10,192 Normal cases, 6,012 instances of Lung Opacity Non-COVID lung infection, and 1,345 Viral Pneumonia images.

 \begin{figure}[hbt!]
 \centering
{\includegraphics[scale=0.4]{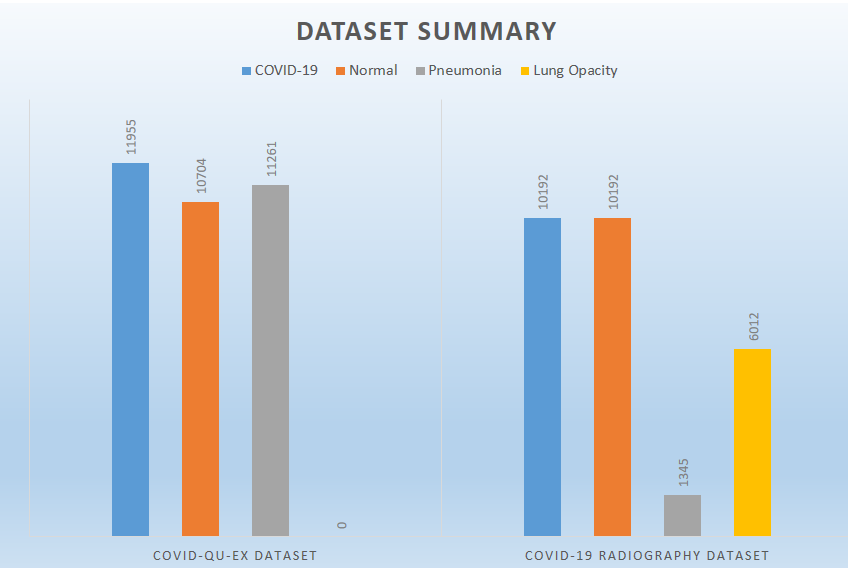}}
  \caption{
  Dataset used for the network}
  \label{diagramm}
\end{figure}

The abnormalities of the sick patient's X-Ray images are well within the lungs or very subtle. So, it does not affect the regular lung shape. We can treat them as regular lung images. These images have corresponding lung masks manually annotated by professional radiologists. The images were resized to $128\times128$ for computational purposes.

\subsection{Experimental Setup}
{

Our entire work was carried out using Python 3.10.12. A high-level API called Keras was used to implement the models; it was constructed on top of the TensorFlow machine learning library.. Our experiments were conducted on google colab platform. The specifications of the Google Colab environment are as follows: a single Tesla K80 GPU with 2496 CUDA cores, accompanied by a single-core hyperthreaded Xeon processor running at 2.3 GHz. The platform includes 13 GB of RAM and 108 GB of runtime HDD. The operating system utilized is based on the Linux Kernel.

}
\subsection{Training Methodology}

\subsubsection{Lung Segmentation-Classification Network }

 The entire lung segmentation and classification network was compiled using specific hyperparameters detailed in Table \ref{table1}. For the segmentation task, the Sigmoid function served as the activation function in the final layer, while the Softmax function was used for the classification module's final layer. We used the Adam optimizer for stochastic gradient descent during model training, which is the combination of the AdaGrad and RMSProp algorithms \cite{adam}. Binary cross-entropy was chosen as the loss function. Binary cross-entropy loss is suitable for segmentation tasks where the goal is to classify each pixel as foreground or background. It compares the predicted probability map to the ground truth binary mask, helping the model accurately delineate objects of interest from the background. Binary cross-entropy loss is  also applicable for multi-label classification tasks, where each sample can have multiple class labels. It handles each class label independently, enabling the model to learn the probability of each label's presence or absence for each sample.

 We can calculate the Binary cross-entropy loss of a prediction $\hat{c}$ and ground truth $c$ using the following equation:
 
\begin{equation}
\label{eq:cce}
    L_{bce} = c \log(\hat{c}) + (1-c) \log(1-\hat{c})
\end{equation}
 
 The batch size for training was set to 32, with an initial learning rate of 0.001. This network was trained using COVID-19 Radiography Dataset\cite{chowdhury2020can}.

\begin{table*}[htbp]
\centering
\caption{Hyperparameters for Segmentation-Classification Module}
\label{table1}
\begin{tabular}{cccc}

\hline
 \multicolumn{2}{c}{\textbf{Lung Segmentation-Classification Network}}& \multicolumn{2}{c}{\textbf{Infection Region Segmentation Network}}\\
 \hline

\textbf{\begin{tabular}[c]{@{}c@{}}Activation function for \\ segmentation final layer\end{tabular}}   & Sigmoid                                                              & \textbf{Activation function in final layer} &Sigmoid                  
\\ 
\textbf{Activation function for Dense layer}                                                           & ReLU                                                                 & - & -\\ 
\textbf{\begin{tabular}[c]{@{}c@{}}Activation function for \\ classification final layer\end{tabular}} & Softmax                                                              & - & - \\ 
\textbf{Optimizer}                                                                                     & Adam                                                                 & \textbf{Optimizer}                                        &Adam                     
\\ 
\textbf{Loss Function}                                                                                 & \begin{tabular}[c]{@{}c@{}}Binary  Crossentropy\end{tabular}& \textbf{Loss Function}                                    &Binary Crossentropy\\ 
\textbf{Batch Size}                                                                                    & 32                                                                   & \textbf{Batch Size}                                       &32                       
\\ 
\textbf{Learning Rate}                                                                                 & 0.001                                                                & \textbf{Learning Rate}                                    &0.001                    \\ \hline
\end{tabular}
\end{table*}

% \begin{table}[htbp]
% \centering
% \caption{Hyperparameters for Segmentation-Classification Module}
% \label{table1}
% \begin{tabular}{|c|c|}
% \hline
% \textbf{\begin{tabular}[c]{@{}c@{}}Activation function for \\ segmentation final layer\end{tabular}}   & Sigmoid                                                             \\ \hline
% \textbf{Activation function for Dense layer}                                                           & ReLU                                                                \\ \hline
% \textbf{\begin{tabular}[c]{@{}c@{}}Activation function for \\ classification final layer\end{tabular}} & Softmax                                                             \\ \hline
% \textbf{Optimizer}                                                                                     & Adam                                                                \\ \hline
% \textbf{Loss Function}                                                                                 & \begin{tabular}[c]{@{}c@{}}Binary  Crossentropy\end{tabular}\\ \hline
% \textbf{Batch Size}                                                                                    & 32                                                                  \\ \hline
% \textbf{Learning Rate}                                                                                 & 0.001                                                               \\ \hline
% \end{tabular}
% \end{table}

\subsubsection{Infection Region Segmentation Network }

The Infection Region Segmentation network was compiled with hyperparameters detailed in Table \ref{table1}. We used the Sigmoid activation function in its final layer and utilized the Adam Optimizer as the optimization algorithm. Categorical cross-entropy used as the loss function for this network. Training was conducted with a batch size of 32 and an initial learning rate set at 0.001. The network was trained using the COVID-QU-Ex Dataset \cite{tahir2021covid}.

% \begin{table}[]
% \centering
% \caption{ Hyperparameters for Infection Module} 
% \label{table2}
% \begin{tabular}{|l|l|}
% \hline
% \textbf{Activation function in final layer} & Sigmoid                  \\ \hline
% \textbf{Optimizer}                                        & Adam                     \\ \hline
% \textbf{Loss Function}                                    & Binary Crossentropy\\ \hline
% \textbf{Batch Size}                                       & 32                       \\ \hline
% \textbf{Learning Rate}                                    & 0.001                    \\ \hline
% \end{tabular}
% \end{table}

The overall training method is shown in algorithm \ref{algoTrainn}.

\begin{algorithm}[h]
\caption{Training Strategy}
{\small
\label{algoTrainn}
\DontPrintSemicolon
  
  \KwInput{Input Chest X-Ray training Images}
  \KwOutput{ {Binary lung masks and Pathology classes from Lung Segmentation-Classification Network. Binary infection masks from Infection Region Segmentation Network.}\\}
  \KwData{\small Utilize pre-trained VGG-16 weights from ImageNet for initializing the first five layers of the proposed models, while the remaining layers are initialized randomly.}

   \BlankLine
\textbf{Step 1 :           Input Initialization} \\
$X_{CXR,train} \gets \textit{CXR images}$ \\
$Y_{lung,train} \gets \textit{Binary lung masks}$ \\
$Y_{label,train} \gets \textit{Pathology classes}$ \\
$Y_{inf,train} \gets \textit{Binary infection masks}$ \\

\textbf{Step 2 :              Fitting models with datasets} \\
$Model_{pipeline} \gets \textit{Lung Segmentation-Classification Network}$ \\
$Model_{infection} \gets \textit{Infection Region Segmentation Network}$ \\

\textbf{Step 3 :              Training models with hyperparameters} \\
\While {$Maximum$ $iteration$ $is$ $not$ $reached$}
{
 \textbf{Train the Lung Segmentation-Classification Network:}
Update the weights of the Lung Segmentation-Classification Network
using the gradient from the Adam optimizer and the loss function 
in the equation \ref{eq:cce}. \\
 $Model_{pipeline}.train(<X_{cxr,train},Y_{lung,train},Y_{label,train}>, Hyperparameters);$ \\ 
 \textbf{Train the Infection Region Segmentation Network:}
Update the weights of the Infection Region Segmentation Network
using the gradient from the Adam optimizer and the loss function 
in the equation \ref{eq:cce}. \\
 $Model_{infection}.train(<X_{cxr,train},Y_{inf,train}>, Hyperparameters);$ 
}}
\end{algorithm}

In model training phase, $Model_{pipeline}$ is trained using the training chest x-ray images $X_{CXR,train}$ and their corresponding binary lung masks $Y_{lung,train}$ and the corresponding disease label $Y_{label,train}$.

The infection region segmentation model, $Model_{infection}$ is trained using the training chest x-ray images $X_{CXR,train}$ and their corresponding binary infection masks $Y_{inf,train}$.

\subsection{Final Output Generation}
\label{sec:final_output}

Algorithm \ref{algoTestt} outlines the process of deriving the final infected region and infection percentage from a chest X-ray image. Given a chest X-ray image $X_{CXR}$ for disease type testing, along with the infected area and infection percentage, where $Model_{pipeline}$ represents the trained model for whole lung segmentation and disease classification, and $Model_{infection}$ is the trained model for infection region segmentation.

In the first place, $Model_{pipeline}$ generates a predicted mask for the lung $Y_{lung}$ and the corresponding label for the predicted disease $Y_{label}$ from the given chest X-ray image. Subsequently, $Model_{infection}$ constructs the probable infected mask for the same chest X-ray image. After thresholding both masks to 0 and 1, the total infected region is calculated using Equation \ref{eq:1}.

The process of generating a final annotated image comprises several steps. Initially, masks in the chest X-ray image are detected, typically represented as binary images indicating the presence or absence of specific objects or regions of interest. Subsequently, the contours of these masks are extracted, which delineate the boundaries of the objects or regions. Finally, the contours of both masks are overlaid onto the original chest X-ray image, resulting in a final annotated image that accentuates the regions of interest. This annotated image serves as a valuable tool for medical 
professionals, providing a clear visual representation of the areas of interest within the chest X-ray, thereby facilitating more accurate and efficient analysis.

\begin{algorithm}[h]
\caption{Strategy of final output generation}
{\small
\label{algoTestt}
\DontPrintSemicolon
  
  \KwInput{{Input chest X-ray image ($X_{CXR}$). Premature binary lung segmentation masks ($Y_{lung}$), disease classification output ($Y_{label}$) from Lung Segmentation-Classification Network. Premature binary infection mask ($Y_{inf}$) from Infection Region Segmentation Network.}\\}
  \KwOutput{{Final predicted disease (fin\_label), percentage of infection (perc), final annotated output (fin\_img)}\\}

   \BlankLine

\For {all pixel value, pix in $Y_{lung}$}
{   
    \If {$pixel$ $value$ $of$ $Y_{lung}$ $is$ $greater$ $than$ $0.6$}
    { 
            $set$ $pixel$ $value$ $of$ $Y_{lung}$ $to$ $1$;
    }
    \Else
    {
           $set$ $pixel$ $value$ $of$ $Y_{lung}$ $to$ $0$;
    }
}

\For {all pixel value, pix in $Y_{inf}$}
{   
    \If {$pixel$ $value$ $of$ $Y_{inf}$ $is$ $greater$ $than$ $0.6$}
    { 
            $set$ $pixel$ $value$ $of$ $Y_{inf}$ $to$ $1$;
    }
    \Else
    {
           $set$ $pixel$ $value$ $of$ $Y_{inf}$ $to$ $0$;
    }
}

\For {$id=1\ldots3$}
    {$fin\_label$ $=$ $find$ $the$ $maximum$ $probability$ $class$ $id$ $from$ $Y_{label}$}

$perc = (\Sigma(Y_{inf})/\Sigma(Y_{lung})) * 100\%$;

$find$ $contour$ $of$ $whole$ $lung$ $from$ $Y_{lung}$; \\
$find$ $contour$ $of$ $infection$ $region$ $from$ $Y_{inf}$;

$fin\_img$ $=$ $superimpose$ $both$ $contours$ $on$ $the$ $X_{CXR}$; 

  }
\end{algorithm}

\subsection{Performance Evaluation Metrics}

The performance of both segmentation modules for the proposed system were evaluated using IoU and Dice Coefficient.\\

The Dice coefficient measures the similarity between predicted and ground truth segmentation masks. Ranging from 0 to 1, where 0 indicates no overlap and 1 indicates perfect overlap, it quantifies segmentation accuracy. Widely used in medical and biological imaging, it helps evaluate the performance of segmentation algorithms.

The Jaccard index, or Jaccard similarity coefficient, is a metric used in image segmentation and object recognition to measure similarity between two sets of data. It quantifies overlap by dividing the size of the intersection of the sets by the size of their union. Scores range from 0 to 1, with higher scores indicating better segmentation accuracy. Score 0 indicates no overlap and 1 indicates perfect overlap. Widely applied in computer vision and pattern recognition, it serves as a key evaluation measure for segmentation algorithms, akin to the Intersection over Union (IoU) metric.

X\textsubscript{1} represents the segmented lung region by the
proposed module, and Y\textsubscript{1} is the ground truth lung region. Then, the dice score is calculated by using the equation \ref{eq:2} and IoU is calculated by using the Eq. \ref{eq:3}.

\begin{equation} \label{eq:2}
    Dice(X,Y) = \frac{2 * \mid{X_1\cap Y_1}\mid}{\mid{X_1}\mid+\mid{Y_1}\mid}
\end{equation}

\begin{equation} \label{eq:3}
    IoU(X,Y) = \frac{\mid{X_1\cap Y_1}\mid}{\mid{X_1}\mid \cup \mid{Y_1}\mid}
\end{equation}

The performance of the classification model was evaluated using sensitivity, precision and accuracy metrics. \\

Sensitivity assesses the ability of a model to accurately detect positive instances from the total number of actual positive instances. It indicates the model's effectiveness in identifying positive cases, measuring the proportion of instances where the model's prediction aligns with the positive class in the ground truth data. Sensitivity is calculated using Eq. \ref{eq:4}.

Precision is an essential measure in machine learning and computer vision to evaluate the ratio of correct positive predictions among all positive predictions generated by a model. It assesses the model's ability to minimize false positives, indicating instances where the model erroneously identifies the positive class while the true ground truth label is negative. Precision is calculated using Eq. \ref{eq:5}.

In the domains of computer vision and machine learning, accuracy is an essential metric since it evaluates the overall correctness of predictions generated by a model. It quantifies the proportion of correct predictions relative to the total number of predictions made. Accuracy can be represented as either a fraction or a percentage. Accuracy is calculated using Eq. \ref{eq:6}

\begin{equation} \label{eq:4}
    Sensitivity = \frac{TP}{TP+FP}
\end{equation}
\begin{equation} \label{eq:5}
    Precision = \frac{TP}{TP+FN}
\end{equation}
\begin{equation} \label{eq:6}
    Accuracy = \frac{TP+TN}{TP+TN+FP+FN}
\end{equation}
\\
where TP, TN, FP and FN denote True Positive, True Negative, False Positive and False Negative. \\

%  ..............
%  Chest InfNet.......End.................
%  ........................

\section{Result and Discussion}
\label{Experimental Result Analysis and Discussion}

%%%%%%%%%%%%%%%%%%%%%%%%%%%%%%%%%%%%%%%%%%   Start    %%%%%%%%%%%%%%%%%%%%%%%%%%%%

%\section{Comparative Result Analysis}

The comparison of the proposed model with four different models using the Chest X-ray dataset was performed to demonstrate its better performance compared to existing models. The four models that were used for comparison were UNet, UNet with transfer learning, Unet++ and Unet++ with
transfer learning. The performance of the proposed model was evaluated and analyzed specifically for lung segmentation, infection region segmentation, and disease classification. Through a comparative analysis between the outcomes of the proposed model and these other models, we demonstrated the effectiveness of the proposed model in terms of its performance in these three areas. These results illustrate that the suggested model has the potential to be an aid for diagnosing patients accurately from chest X-rays.

\subsection{Comparison on Evaluation Metrics}

Our findings indicate that the proposed networks outperform most existing state-of-the-art models in chest X-ray datasets. We assessed both the Lung Segmentation and Disease Classification Network and the Infection Region Segmentation Network using suitable evaluation metrics.

For the Lung Segmentation and Disease Classification Network, which performs both segmentation and classification tasks, we evaluated its performance using precision, recall, and accuracy for classification, and metrics such as the dice coefficient and Jaccard index for segmentation. Conversely, as the Infection Region Segmentation Model solely focuses on segmenting infected areas, we evaluated its performance using the dice coefficient and Jaccard index metrics.

\subsubsection{Lung Segmentation-Classification Network}

%%%%%%%%%%%%%%%%%%%%%%%%%%%%%%

We found out that the proposed Lung Segmentation and Disease Classification network performs better than most existing state-of-the-art models. For the Lung Segmentation and Disease Classification network, the whole lung segmentation task was evaluated with the dice coefficient and Jaccard index scores, and the disease classification task was evaluated using precision, recall, and accuracy metrics.

 As the proposed Lung Segmentation and Disease Classification network has the ability to perform two tasks, the first task is to segment out the lung portion from the chest x-ray. On the COVID-19 Radiography Dataset, this proposed network gained the top position obtaining the dice score of 97.61\% and the Jaccard index of 93.59\%. We evaluated and compared the following dataset with four other different models, which are: U-Net without transfer learning, U-Net with transfer learning, U-Net++ without transfer learning, and U-Net++ with transfer learning. Dice scores for these models are 94.20, 95.41, 95.65 and 97.63, respectively. The Jaccard Index scores for these models are 89.63, 91.60,91.73 and 93.47, respectively. By evaluating the relative performances with the following models, we can see that our proposed network obtained the highest Jaccard Index among these models but U-Net++ with transfer learning obtained the highest dice coefficient score as it contains a larger number of parameters than the proposed network. But as a lightweight model, our proposed network performs almost better than the other models for the segmentation task.

Comparison between different models for the segmentation task using the COVID-19 Radiography Dataset is summarized in Table \ref{segclass}.

\begin{table}
    \centering
    \caption{Comparison of different Architectures for Lung Segmentation}
    \label{segclass}
    \begin{adjustbox}{width=\columnwidth,center}
    \Large
\begin{tabular}{cccccc}
\hline
\textbf{Comparison} & \textbf{UNET} & \textbf{UNET(TL)} & \textbf{UNET++} & \textbf{UNET++(TL)} & \textbf{Proposed Model} \\ \hline
\textbf{IoU (\%)}    & 89.63         & 91.60             & 91.73           & 93.47               & \textbf{93.59}          \\ 
\textbf{Dice (\%)}  & 94.20         & 95.41             & 95.65           & \textbf{97.63}      & 97.61                   \\ \hline
\end{tabular}
\end{adjustbox}
\end{table}

\begin{figure}[!ht]
    \centering
    \includegraphics[width = \columnwidth]{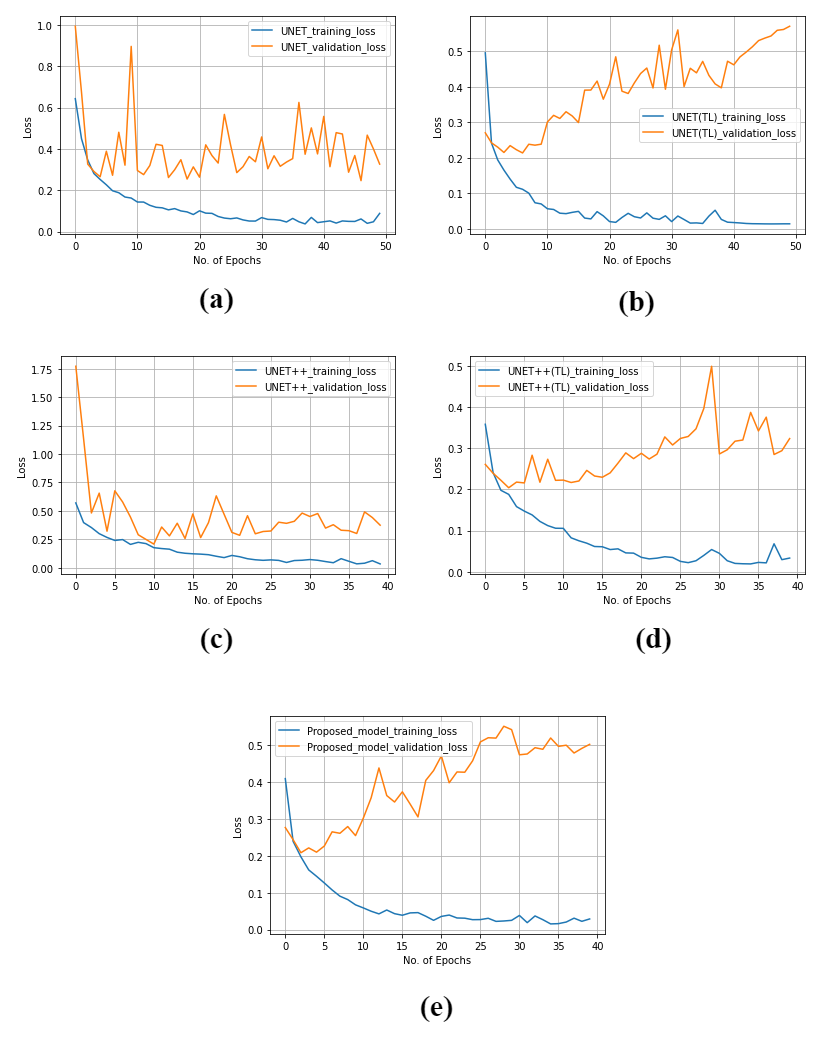}
    \caption{Progression graph of the loss function over the number of epochs on the models (a) U-Net without transfer learning, (b) U-Net with transfer learning, (c) U-Net++ without transfer learning, (d) U-Net++ with transfer learning, and (e) Proposed Model.}
    \label{graph1}
\end{figure}

\begin{figure}[t]
    \centering
    \includegraphics[width = \columnwidth]{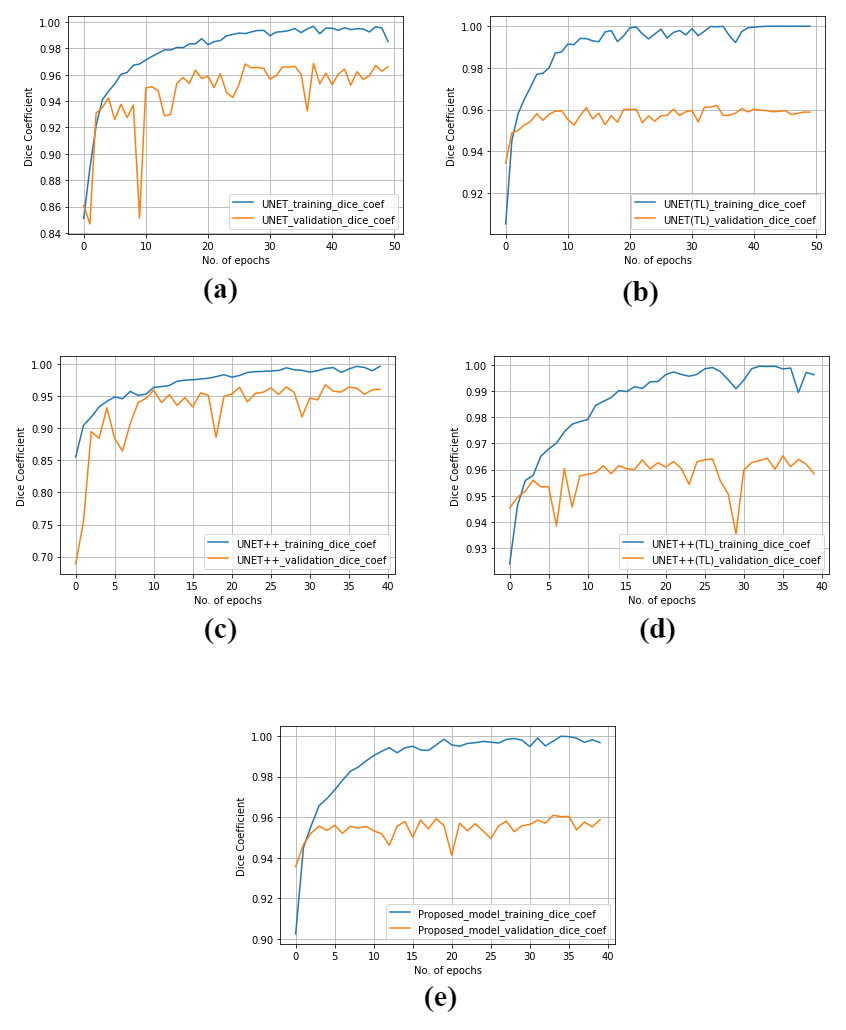}
    \caption{Progression graph of the metric dice coefficient over the number of epochs on the models (a) U-Net without transfer learning, (b) U-Net with transfer learning, (c) U-Net++ without transfer learning, (d) U-Net++ with transfer learning, and (e) Proposed Model.}
    \label{graph2}
\end{figure}

The efficient learning ability of our proposed network is proved by considering the progress of performance with epochs. We have run each of the models for 50 epochs. From our experiment, we observed that our proposed model converges much faster than U-Net and similar models also provide a better result comparatively. These results suggest that using our proposed network, we can achieve a superior result in a fewer number of training epochs than the traditional U-Net architecture. This fact is shown in Figure \ref{graph1}, \ref{graph2} in terms of the progression of the loss function and dice coefficient. It clearly shows that adding the transfer learning technique helped the proposed network to converge faster in all cases.

% \begin{figure}[t]
%     \centering
%     \includegraphics[scale=.4]{class_seg Plot/jaccard.pdf}
%     \caption{Progression graph of the metric Jaccard index over the number of epochs on the models (a) U-Net without transfer learning, (b) U-Net with transfer learning, (c) U-Net++ without transfer learning, (d) U-Net++ with transfer learning, and (e) Proposed Model.}
%     \label{graph3}
% \end{figure}

The second task of the proposed Lung Segmentation and Disease Classification network is to classify the type of disease contained by the chest x-ray. On the COVID-19 Radiography Dataset, this proposed model gained the top position obtaining an accuracy of 93.86\% comparing with U-Net without transfer learning, U-Net with transfer learning, U-Net++ without transfer learning, and U-Net++ with transfer learning. These models gained accuracy 91.70\%, 92,79\%, 90.48\% and 92.06\% respectively. Precision for these models are 91.09, 89.18, 91.12 and 89.86, respectively and our proposed network scored 89.75 which is almost similar. Recall for these models are 88.97, 89.11, 89.64 and 89.79, respectively and our proposed network scored 89.55 which is almost similar. By evaluating the relative performances with the following models, we can see that our proposed network obtained the highest accuracy among these models but U-Net++ with transfer learning obtained the highest recall as it contains a large number of parameters than the proposed network and U-Net without transfer learning obtained the highest precision as it contains more convolution layers which can extract better features than the proposed network. But as a lightweight model, our proposed network performs almost better than the other models for the classification task as well.

\begin{figure}[t]
    \centering
    \includegraphics[width = \columnwidth]{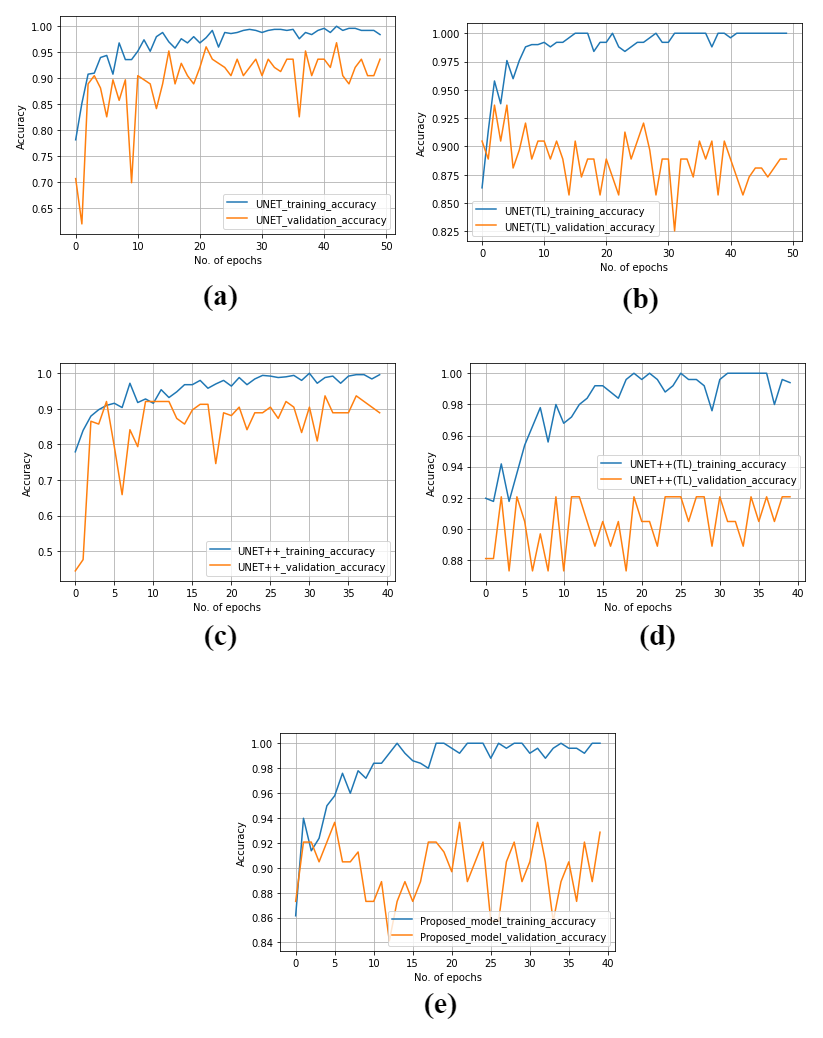}
    \caption{Progression graph of the metric accuracy over the number of epochs on the models for classification task (a) U-Net without transfer learning, (b) U-Net with transfer learning, (c) U-Net++ without transfer learning, (d) U-Net++ with transfer learning, and (e) Proposed Model.}
    \label{graph4}
\end{figure}

\begin{figure}[t]
    \centering
    \includegraphics[width = \columnwidth]{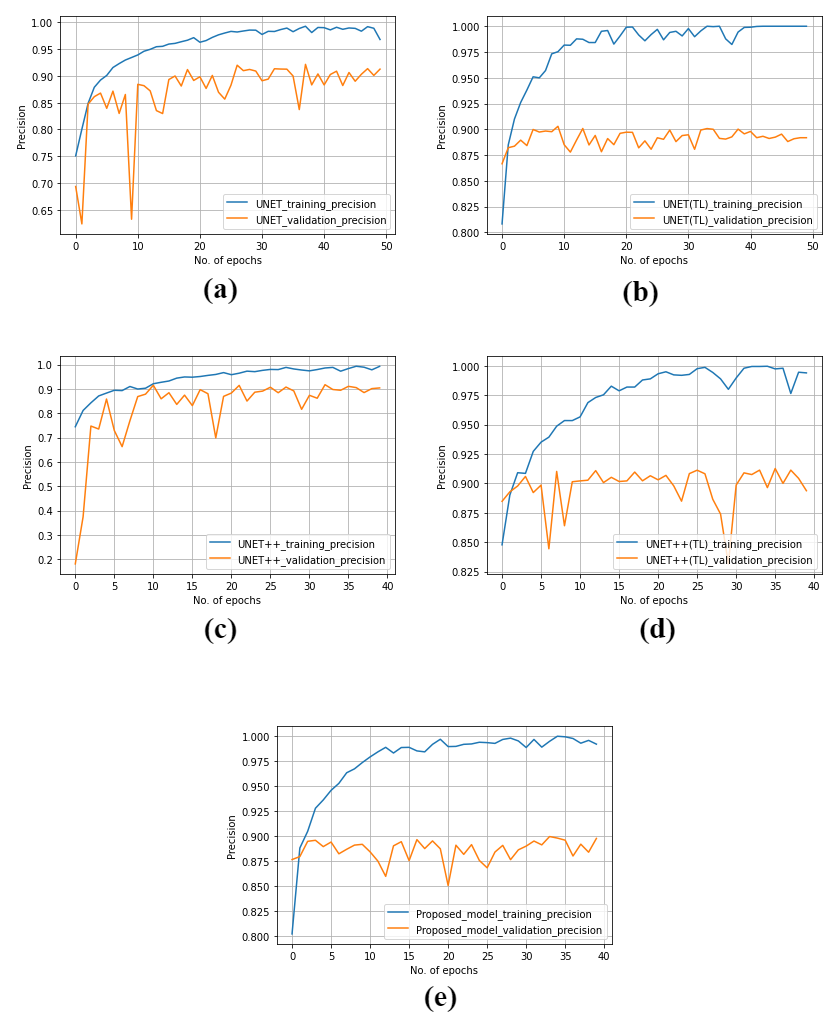}
    \caption{Progression graph of the metric precision over the number of epochs on the models for classification task (a) U-Net without transfer learning, (b) U-Net with transfer learning, (c) U-Net++ without transfer learning, (d) U-Net++ with transfer learning, and (e) Proposed Model.}
    \label{graph5}
\end{figure}

Comparison between different models for the classification task using the COVID-19 Radiography Dataset is summarized in Table \ref{segpred}. Figure \ref{graph4}, \ref{graph5} shows the progression of the accuracy and precision of different U-Net variants by fusing our classification module. It clearly shows that our proposed network performs better and converges faster in all cases.

\begin{table}
    \centering
    \caption{Comparison of different Architectures for Diseases Classification}
    \label{segpred}
    \begin{adjustbox}{width=\columnwidth,center}
    \Large
\begin{tabular}{cccccc}
\hline
\textbf{Comparison}    & \textbf{UNET} & \textbf{UNET(TL)} & \textbf{UNET++} & \textbf{UNET++(TL)} & \textbf{Proposed Model} \\ \hline
\textbf{Precision(\%)} & 91.09& 89.18             & \textbf{91.12}& 89.86               & 89.75                   \\ 
\textbf{Recall(\%)}    & 88.97         & 89.11             & 89.64           & \textbf{89.79}      & 89.55                   \\ 
\textbf{Accuracy (\%)} & 91.70         & 92.79             & 90.48           & 92.06               & \textbf{93.86}          \\ \hline
\end{tabular}
\end{adjustbox}
\end{table}

% \begin{figure}[t]
%     \centering
%     \includegraphics[scale=.4]{class_seg Plot/recall.pdf}
%     \caption{Progression graph of the metric recall over the number of epochs on the models for classification task (a) U-Net without transfer learning, (b) U-Net with transfer learning, (c) U-Net++ without transfer learning, (d) U-Net++ with transfer learning, and (e) Proposed Model.}
%     \label{graph6}
% \end{figure}

\begin{figure}[t]
    \centering
    \includegraphics[width=\columnwidth]{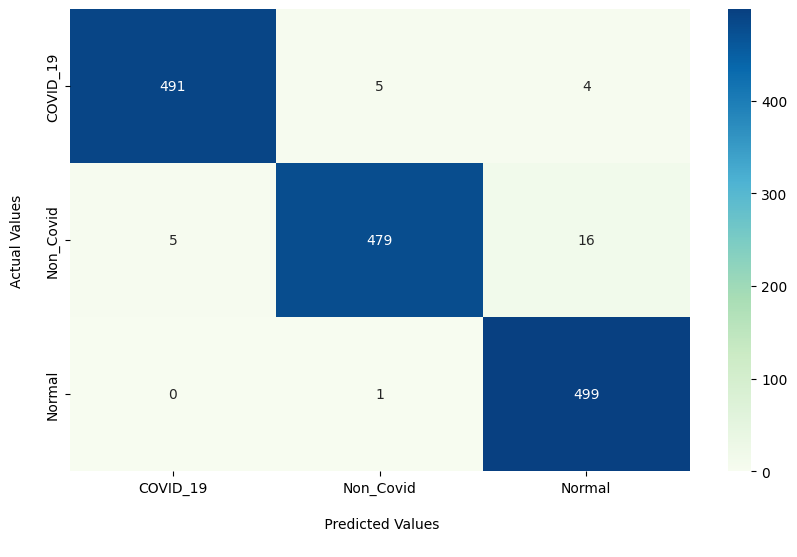}
    \caption{Confusion matrix of the proposed model for classification task.}
    \label{conf}
\end{figure}

Figure \ref{conf} represents the confusion matrix for classifying lung diseases for three categories which are Covid-19, Lung Opacity, and Normal chest x-rays respectively.
We tested our network with 1500 chest X-ray, 500 X-ray for each class. From 500 Covid-19 affected Chest X-ray, our network predicted 491 X-ray correctly and 9 X-ray were misclassified. Also from 500 viral pneumonia affected Chest X-ray, our network predicted 479 X-ray correctly and 21 X-ray were misclassified and from 500 normal Chest X-ray, our network predicted 499 X-ray correctly and 1 X-ray were misclassified.

Table \ref{table3} shows that our proposed network performs better than most of the existing models using this dataset.

\begin{table}[]
\centering
\caption{{Comparison of proposed Segmentation-Classification Model with existing works} }
\label{table3}
\begin{tabular}{cccc}
\hline
\textbf{Research Paper} & \textbf{\begin{tabular}[c]{@{}c@{}}Segmentation\\  DSC\end{tabular}} & \textbf{Accuracy} & \textbf{\begin{tabular}[c]{@{}c@{}}Classification \\ Sensitivity\end{tabular}} \\ \hline
Yeh et al. {\cite{yeh2020cascaded}}     & 0.88                                                                 & -                 & 82\%                                                                           \\ 
Tabik et al. {\cite{tabik2020covidgr}}   & 0.885                                                                & 76\%              & 73\%                                                                           \\ 
Robert et al.{\cite{hertel2022deep}}        & 0.95                                                                 & 84\%              & 82\%                                                                           \\ 
\textbf{Proposed Model}          & \textbf{0.97}                                                                 & \textbf{93.86}\%           & \textbf{89.75}\%                                                                        \\ \hline
\end{tabular}
\end{table}

%%%%%%%%%%%%%%%%%%%%%%%%%%

\subsubsection{Infection Region Segmentation Network}

%%%%%%%%%%%%%%%%%%%%%%%%%

We found that the Infection Region Segmentation Netwo we proposed in our study was quite effective in identifying the lung areas that were infected from chest X-ray images.

We evaluated our proposed network on the COVID-QU-Ex Dataset and our model achieved comparatively better results, obtaining a Dice score of 87.61\%, Jaccard Index score of 97.67\% and an accuracy of 98.23\%. We conducted comparative evaluations with four other models, which are: U-Net without transfer learning, U-Net with transfer learning, U-Net++ without transfer learning, and U-Net++ with transfer learning. Dice scores for these models are 73.50, 76.52, 77.68 and 84.94 respectively. The Jaccard Index scores for these models are 93.67, 95.42, 96.41 and 97.70, respectively. And finally the Accuracy for these models are 94.53, 96.35, 95.19 and 96.03 respectively. By evaluating the relative performances with the following models, we can see that our proposed network obtained the highest Dice score and Accuracy among these models but U-Net++ with transfer learning obtained the highest Jaccard Index score as it contains a larger number of parameters than the proposed network. Despite its lightweight design, our proposed network outperformed the other models in terms of both dice coefficient and accuracy.

\begin{table}
    \centering
    \caption{Comparison of different Architectures for Infection Localization}
    \label{lala}
    \begin{adjustbox}{width=\columnwidth,center}
    \Large
\begin{tabular}{cccccc}
\hline
\textbf{Comparison}    & \textbf{UNET} & \textbf{UNET (TL)} & \textbf{UNET++} & \textbf{UNET++ (TL)} & \textbf{Proposed Infection Model} \\ \hline
\textbf{IoU(\%)}       & 93.67         & 95.42              & 96.41           & \textbf{97.70}       & 97.67                    \\ 
\textbf{Dice (\%)}     & 73.50         & 76.52              & 77.68           & 84.94       & \textbf{87.61}                    \\ 
\textbf{Accuracy (\%)} & 94.53         & 96.35              & 95.19           & 96.03                & \textbf{98.23}                    \\ \hline
\end{tabular}
\end{adjustbox}
\end{table}

\begin{figure}[t]
    \centering
    \includegraphics[width = \columnwidth]{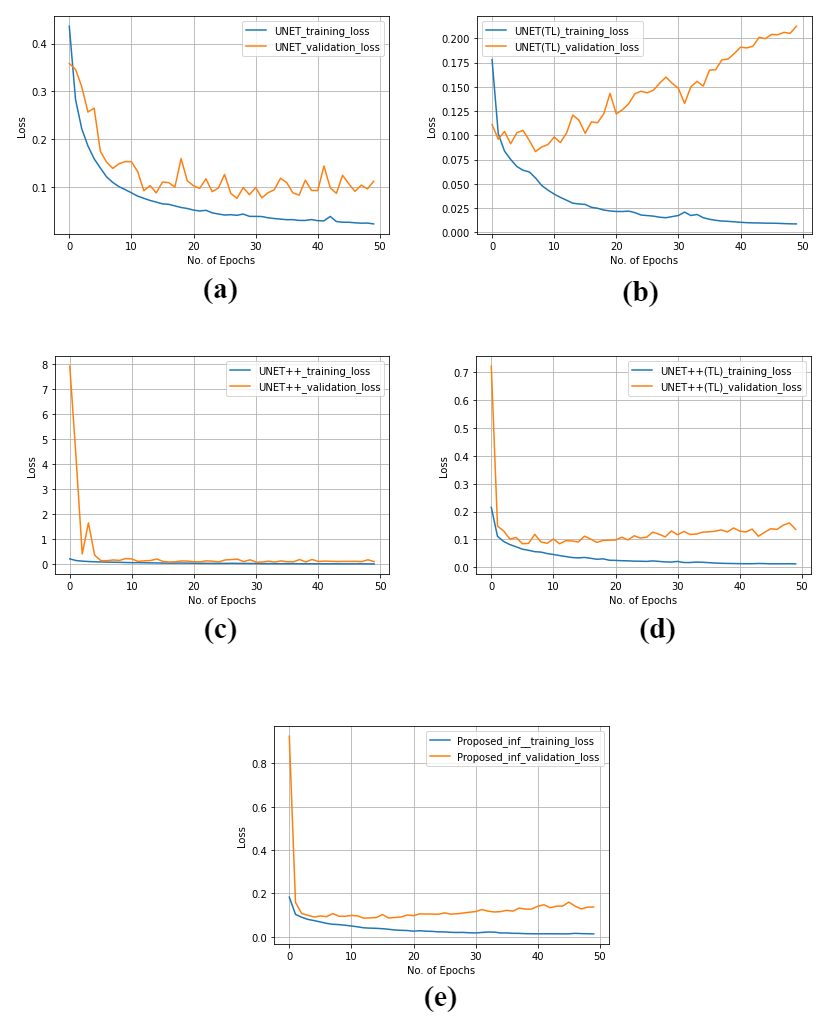}
    \caption{Progression graph of the loss function over the number of epochs on the infection models (a) U-Net without transfer learning, (b) U-Net with transfer learning, (c) U-Net++ without transfer learning, (d) U-Net++ with transfer learning, and (e) Proposed Model.}
    \label{graph_inf1}
\end{figure}

\begin{figure}[t]
    \centering
    \includegraphics[width = \columnwidth]{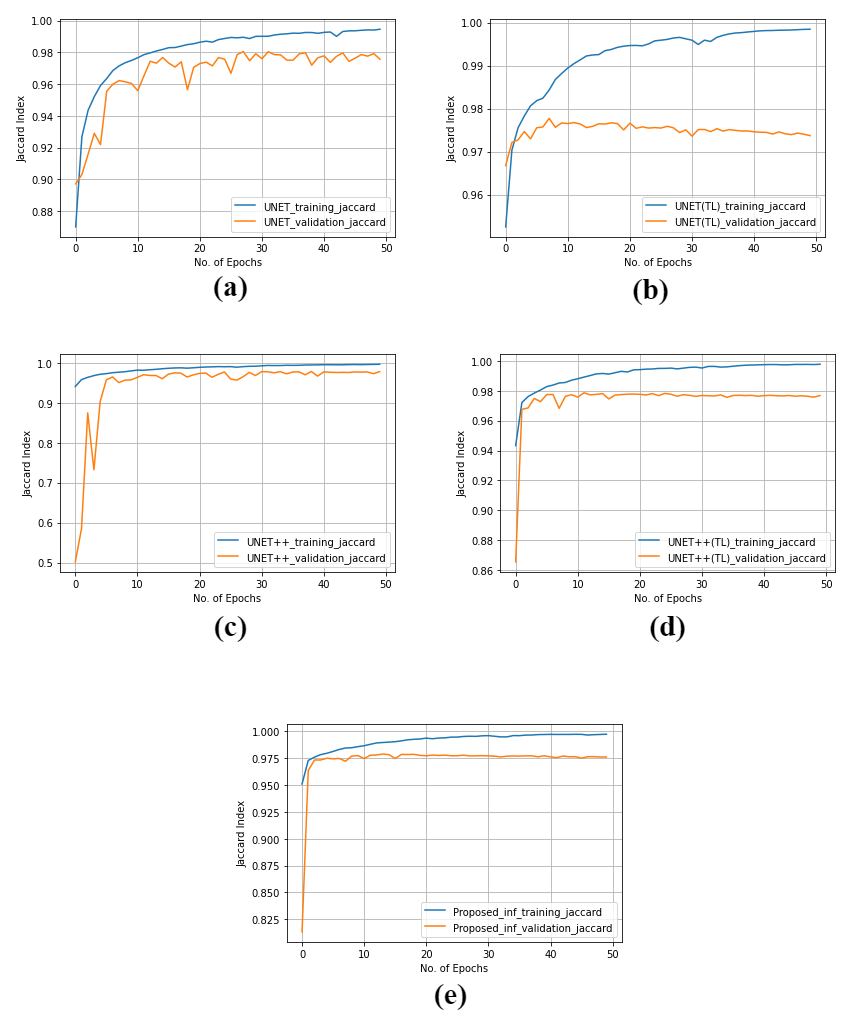}
    \caption{Progression graph of the metric Jaccard index over the number of epochs on the infection models (a) U-Net without transfer learning, (b) U-Net with transfer learning, (c) U-Net++ without transfer learning, (d) U-Net++ with transfer learning, and (e) Proposed Model.}
    \label{graph_inf3}
\end{figure}

The comparison between different models for the segmentation task, as performed on the COVID-QU-Ex Dataset, is summarized in Table \ref{lala}. Our proposed network consistently demonstrated superior performance metrics compared to the other models, highlighting its efficacy in segmenting infected regions from chest X-ray images. Furthermore, our experiments revealed the efficiency of our proposed model in terms of learning capabilities over a fixed number of epochs. By running experiments for 50 epochs, we observed that our model converged faster than traditional U-Net architectures while performing better  results. This improvement was evident across various evaluation metrics, including the loss function and Jaccard index, as depicted in Figure \ref{graph_inf1} and \ref{graph_inf3}.

% \begin{figure}[t]
%     \centering
%     \includegraphics[scale=.4]{Infection Plot/dice.pdf}
%     \caption{Progression graph of the metric dice coefficient over the number of epochs on the infection models (a) U-Net without transfer learning, (b) U-Net with transfer learning, (c) U-Net++ without transfer learning, (d) U-Net++ with transfer learning, and (e) Proposed Model.}
%     \label{graph_inf2}
% \end{figure}

\begin{figure*}[!ht] %!t
    \centering
    \includegraphics[width=\textwidth,height=20cm]{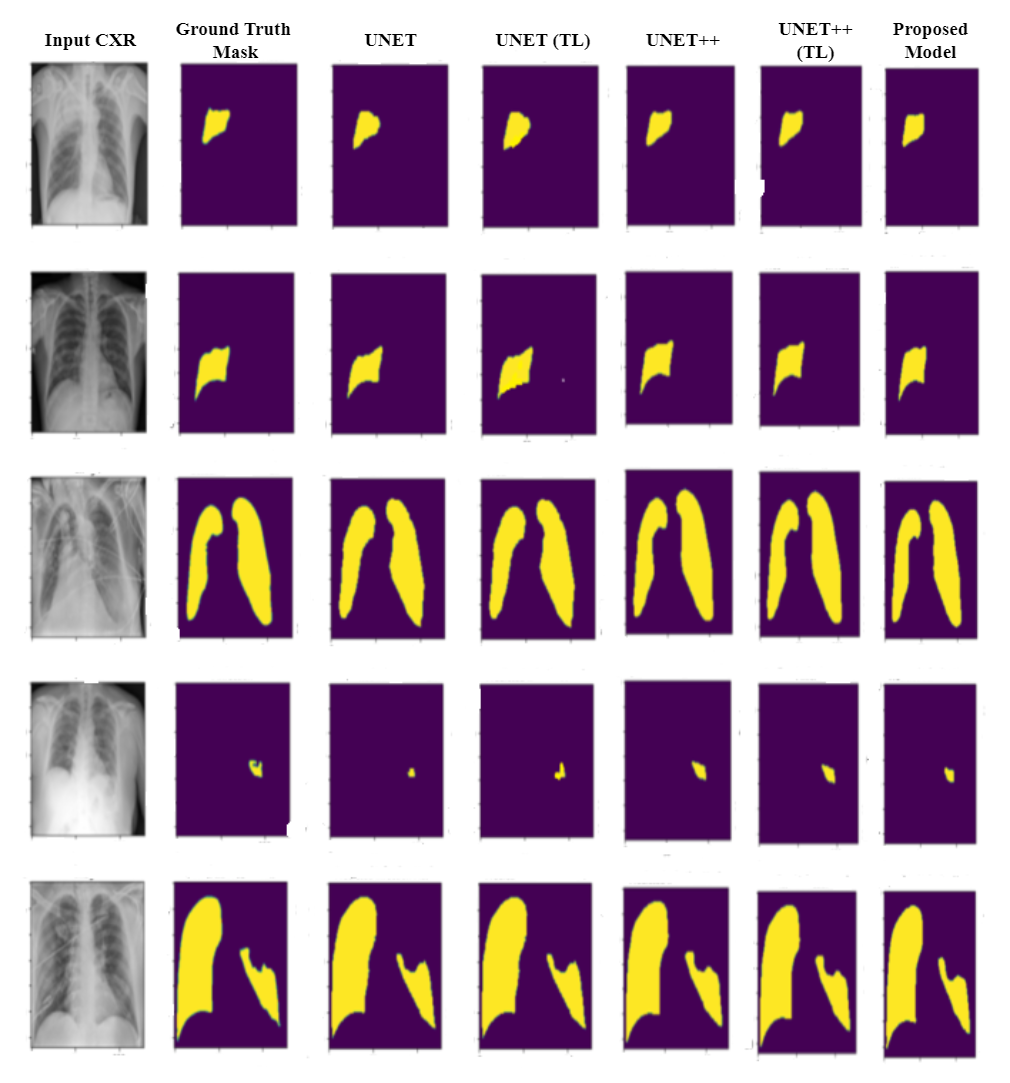}
    \caption{Comparison of Infection Region Segmentation Model with various segmentation models}
    \label{final_inf}
\end{figure*}

A comparative analysis of outputs for different models is illustrated in Figure \ref{final_inf}, further emphasizing the superiority of our proposed infection segmentation model. Table \ref{inftable} summarizes the comparative performance of various models on the COVID-QU-Ex Dataset, with our proposed network consistently outperforming the others. Our findings emphasize the efficacy and efficiency of the proposed Infection Region Segmentation Network overall, indicating that it is a potentially useful tool for precisely locating infected regions in lung X-ray images.

%%%%%%%%%%%%%%%%%%%%%%%%%

\begin{table}
\centering
\caption{{Comparison of proposed Infection Model with existing works}} 
\label{inftable}
\begin{tabular}{cccc}
\hline
\textbf{Research Paper} & \textbf{Accuracy} & \textbf{DSC}     & \textbf{IoU}     \\ \hline
Anas et al. {\cite{tahir2021covid}}    & 97.99\%           & \textbf{88.1}\%           & 83.1\%           \\ 
Covid-MANet {\cite{sharma2022covid}}    & 97.01\%           & 86.15\%          & 76.94\%          \\
\textbf{Proposed module} & \textbf{98.23\%}  & 87.61\% & \textbf{97.67\%} \\ \hline
\end{tabular}
\end{table}

\begin{table*}

    \centering
    \caption{Total No. of Parameters and Average Time for different Architectures}
    \label{papa}
    \begin{adjustbox}{width=\columnwidth,center}
    \Large
\begin{tabular}{cccccc}

\hline
 \multicolumn{3}{c}{\textbf{Lung Segmentation-Classification Network}}& \multicolumn{3}{c}{\textbf{Infection Region Segmentation Network}}\\
\hline
\textbf{Model with Classification Module} & \textbf{No. of Parameters} & \textbf{Average Training time per epoch (Second)}  & \textbf{Infection localization model} & \textbf{No. of Parameters} &\textbf{Average Training time per epoch (Second)} 
\\ \hline
\textbf{UNet}                             & 29,837,267                    & 113                                                 & \textbf{UNet}                             & 27,887,169                    &91                                                
\\ 
\textbf{UNet with transfer learning}      & 21,821,637                    & 101                                                 & \textbf{UNet with transfer learning}      & 19,854,657                    &76                                                
\\ 
\textbf{UNet++}                           & 42,970,501                    & 287                                                & \textbf{UNet++}                           & 36,644,225                    &124                                               
\\ 
\textbf{UNet++ with transfer learning}    & 34,107,525                    & 269                                                & \textbf{UNet++ with transfer learning}    & 30,140,545                    &105                                               
\\ 
\textbf{Proposed Model}                   & 28,132,805                    & 107                                                 & \textbf{Proposed Model}                   & 26,165,825                    &83                                                \\ \hline
\end{tabular}
\end{adjustbox}
\end{table*}

\subsection{Evaluation of the Severity of Infection}

In our proposed system, the generation of semantic segmented lung masks and segmented infection region masks involves two separate models: the Segmentation-Classification model and the Infection Region Segmentation model, respectively. The Segmentation-Classification model is responsible for constructing the entire semantic segmented lung mask, which outlines the boundaries of the lungs in the Chest X-ray image. Conversely, the Infection Region Segmentation model generates the segmented infection region mask, delineating the boundaries of the infected area within the lungs. By overlaying these two masks, the system can determine the infection area in the lungs by extracting the borders of both masks. This process enables the system to precisely locate the infection within the lungs. Subsequently, the extracted areas from both segmented masks are superimposed onto the original Chest X-ray image. This results in a visual depiction of the infection area within the lungs, providing medical professionals with valuable insights for accurate diagnoses and treatment planning. Compared to traditional methods, this approach offers a more effective and efficient means of analyzing Chest X-rays.

\begin{figure*}[!ht] %!t
    \centering
    \includegraphics[width=\textwidth, height = 15cm]{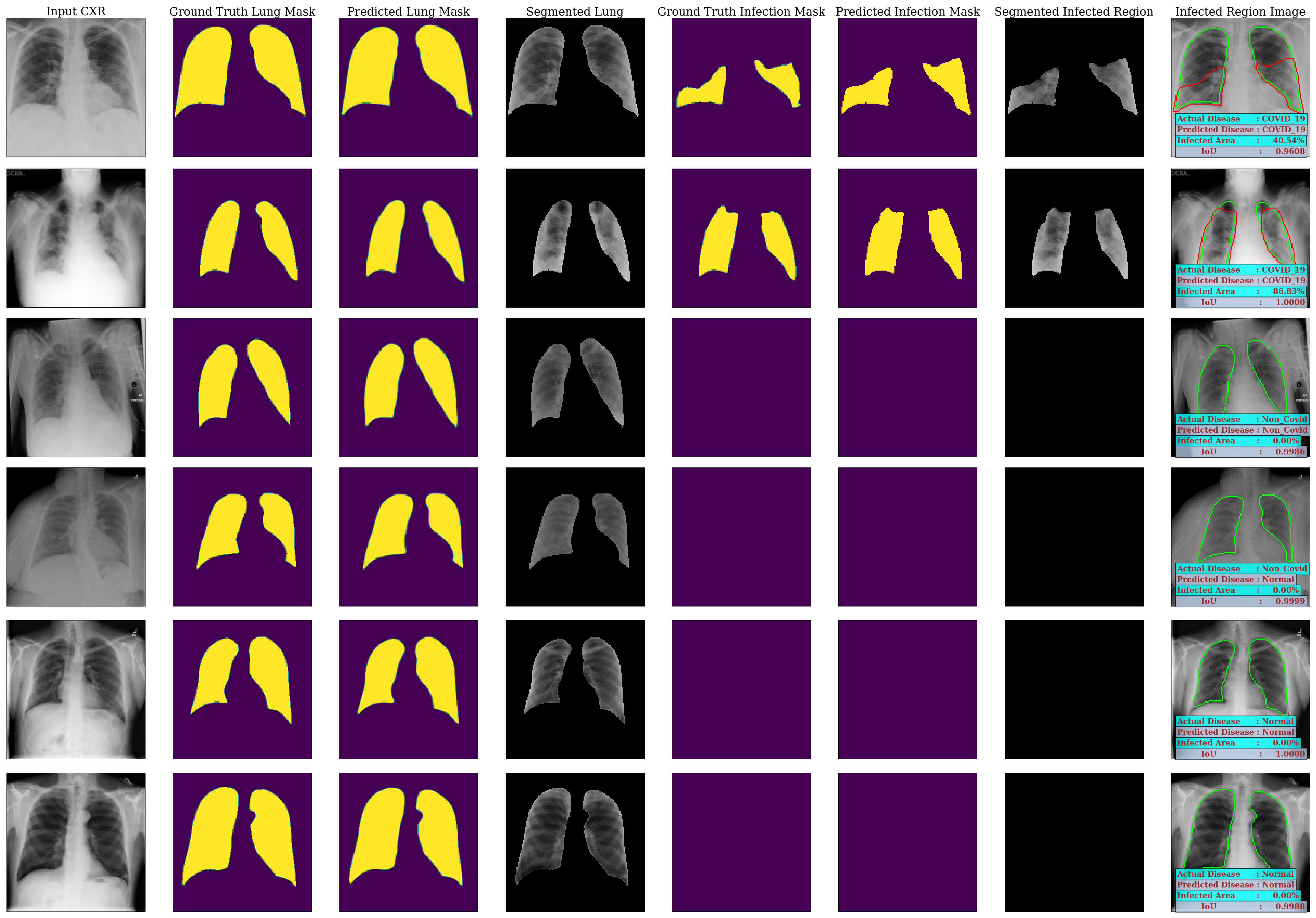}
    \caption{Infection Region Localization and Evaluation of the Proposed System}
    \label{final}
\end{figure*}

The severity of the Infection was evaluated using Eq. \ref{eq:1} and the exactness of the infection localization was evaluated using Eq. \ref{eq:8}. The infection region localization and severity as well as correctness evaluation of our proposed system is illustrated in Fig. \ref{final}.

\subsection{Proposed System is Lightweight}
\label{res:param}

As previously mentioned, our proposed Lung Segmentation and Disease Classification network contains fewer parameters compared to the most state-of-the-art networks. Table \ref{papa} illustrates this comparison between the parameter counts of our proposed architecture and four other architectures including U-Net without transfer learning, U-Net with transfer learning, U-Net++ without transfer learning, and U-Net++ with transfer learning. Our proposed network contains 28M parameters including the classification module and takes 107 seconds for training in every epoch. Other models contain 30M, 22M, 43M and 34M respectively and takes 113, 101, 287 and 269 seconds respectively, for training in every epoch. Notably, our proposed architecture stands out for its efficiency, requiring less time for execution compared to its counterparts.

% \begin{table}

%     \centering
%     \caption{Total No. of Parameters and Average Time for different Architectures}
%     \label{papa}
%     \begin{adjustbox}{width=\columnwidth,center}
% \begin{tabular}{|c|c|c|}
% \hline
% \textbf{Model with Classification Module} & \textbf{No. of Parameters} & \textbf{Average Training time per epoch (Second)} \\ \hline
% \textbf{UNet}                             & 29,837,267                    & 113                                                \\ \hline
% \textbf{UNet with transfer learning}      & 21,821,637                    & 101                                                \\ \hline
% \textbf{UNet++}                           & 42,970,501                    & 287                                               \\ \hline
% \textbf{UNet++ with transfer learning}    & 34,107,525                    & 269                                               \\ \hline
% \textbf{Proposed Model}                   & 28,132,805                    & 107                                                \\ \hline
% \end{tabular}
% \end{adjustbox}
% \end{table}

Similarly, the proposed Infection Region Segmentation network also features fewer parameters compared to leading networks. Table \ref{papa} presents a comparison of the parameter counts between our proposed model and others. Our proposed network contains 26M parameters without the classification module and takes 83 seconds for training in every epoch. Other models contain 28M, 20M, 36.5M and 30M respectively and takes 91, 76, 124 and 105 seconds respectively, for training in every epoch. We can observe that as the number of trainable parameters decreases, the time required to complete one epoch also decreases. This highlights the efficiency and effectiveness of our proposed system in terms of parameter optimization and computational resource utilization.

% \begin{table}

%     \centering
%     \caption{Total No. of Parameters and Average Time for different Architectures}
%     \label{jaja}
%     \begin{adjustbox}{width=\columnwidth,center}
% \begin{tabular}{|c|c|c|}
% \hline
% \textbf{Infection localization model} & \textbf{No. of Parameters} & \textbf{Average Training time per epoch (Second)} \\ \hline
% \textbf{UNet}                             & 27,887,169                    & 91                                                \\ \hline
% \textbf{UNet with transfer learning}      & 19,854,657                    & 76                                                \\ \hline
% \textbf{UNet++}                           & 36,644,225                    & 124                                               \\ \hline
% \textbf{UNet++ with transfer learning}    & 30,140,545                    & 105                                               \\ \hline
% \textbf{Proposed Model}                   & 26,165,825                    & 83                                                \\ \hline
% \end{tabular}
% \end{adjustbox}
% \end{table}

%%%%%%%%%%%%%%%%%%%%%%%%%%%%%%%%%%%%%%%   End  %%%%%%%%%%%%%%%%%%%%%%%%%%%%%%%%%%%%%%

%%%%%%%%%%%%%%%%%%%%%%%%%%%%%%%%%%%%%%%%%%%%%%%%%%%%%%%%%%%%%%%%%%

\section{Conclusion}

This work presents a significant advancement in medical imaging, particularly for detecting and localizing lung infections using chest X-ray (CXR) images. The integration of a novel, lightweight deep-learning-based segmentation-classification network addresses several challenges in diagnosing lung diseases such as COVID-19 and pneumonia. Leveraging transfer learning with pre-trained VGG-16 weights, the model performs robustly with limited training data. Incorporating refined skip connections within the UNet++ framework enhances segmentation precision by reducing semantic gaps. Additionally, the classification module at the end of the encoder block enables simultaneous classification and segmentation, enhancing versatility and providing comprehensive diagnostic insights.

Experimental results demonstrate the model's superiority in terms of accuracy and computational efficiency compared to existing methods. Achieving an Intersection over Union (IoU) of 93.59\% and a Dice Similarity Coefficient (DSC) of 97.61\% for lung area segmentation, along with an IoU of 97.67\% and a DSC of 87.61\% for infection region localization, highlights the model's efficacy. Furthermore, the model's high accuracy of 93.86\% and sensitivity of 89.55\% in detecting chest diseases confirm its reliability and practical applicability.

The streamlined and lightweight design facilitates easier hyperparameter tuning and deployment on edge devices, making it suitable for real-time and resource-constrained environments. This work broadens the applicability of advanced deep learning architectures in medical image analysis and underscores their potential to significantly improve clinical outcomes through precise, efficient, and comprehensive diagnostic solutions. Future research may focus on further optimizing this model and extending its application to other types of medical imaging and diseases, thereby enhancing its utility in diverse healthcare settings.

\bibliographystyle{model6-num-names}
\bibliography{Bibliography}
\end{document}